\documentclass[runningheads]{llncs}
\usepackage{graphicx}
\usepackage[noend]{algpseudocode}
\usepackage{algorithm,algorithmicx}
\usepackage{amsmath}
\usepackage{amssymb}
\usepackage{hyperref}
\usepackage{rotating}
\usepackage{multirow}
\usepackage{todonotes}
\usepackage{graphicx}
\usepackage[export]{adjustbox}
\usepackage{xcolor}
\usepackage{caption}
\usepackage{wrapfig}
\usepackage{picins}
\usepackage{bm}

\begin{document}
\title{Towards learned optimal $q$-space sampling in diffusion MRI}

\newcommand{\bb}[1]{\bm{\mathrm{#1}}}
\newcommand{\Tr}{\mathrm{T}}
\newcommand{\uul}[1]{{\bf #1}}
%

\author{Tomer Weiss\inst{1} \and
Sanketh Vedula\inst{1} \and
Ortal Senouf\inst{1} \and
Oleg Michailovich\inst{2} \and
Alex Bronstein\inst{1}}
\authorrunning{T. Weiss et al.}
%
\institute{Technion, Israel. \and
University of Waterloo, Canada.}

\maketitle              
\begin{abstract}
Fiber tractography is an important tool of computational neuroscience that enables reconstructing the spatial connectivity and organization of white matter of the brain. Fiber tractography takes advantage of diffusion Magnetic Resonance Imaging (dMRI) which allows measuring the apparent diffusivity of cerebral water along different spatial directions. Unfortunately, collecting such data comes at the price of reduced spatial resolution and substantially elevated acquisition times, which limits the clinical applicability of dMRI. This problem has been thus far addressed using two principal strategies. Most of the efforts have been extended towards improving the quality of signal estimation for any, yet {\it fixed} sampling scheme (defined through the choice of diffusion-encoding gradients). On the other hand, optimization over the sampling scheme has also proven to be effective. Inspired by the previous results, the present work consolidates the above strategies into a unified estimation framework, in which the optimization is carried out with respect to both estimation model and sampling design {\it concurrently}. The proposed solution offers substantial improvements in the quality of signal estimation as well as the accuracy of ensuing analysis by means of fiber tractography. While proving the optimality of the learned estimation models would probably need more extensive evaluation, we nevertheless claim that the learned sampling schemes can be of immediate use, offering a way to improve the dMRI analysis without the necessity of deploying the neural network used for their estimation. We present a comprehensive comparative analysis based on the Human Connectome Project data. \\
Code and learned sampling designs available at \url{https://github.com/tomer196/Learned_dMRI}.
\end{abstract}
\section{Introduction}
Fiber tractography\index{Fiber tractography} has long become a standard tool of computational neuroscience which makes it possible to delineate the structure of neural fiber tracts within white matter, thus facilitating quantitative assessment of its integrity and connectivity in application to clinical diagnosis \cite{Bullmore2009ComplexBN,duffau2011brain}. 

The accuracy of fiber tractography, however, depends on the quality of diffusion MRI (dMRI) data used for estimation of the local directions of fiber tracts at each spatial voxel. Such data is usually available as a collection of 3D MRI volumes, known as diffusion-encoded images, which represent signal attenuation due to water diffusion along various directions and different levels of diffusion sensitization. In particular, high angular resolution diffusion imaging (HARDI) \cite{HARDI2008}\index{HARDI}, were each diffusion-encoding image representing a single sampling point on the spherical shell. In this case, collection of a relatively large number of samples (as it is often required by more advanced methods of diffusion data analysis) entails longer imaging sessions, which tend to be avoided in clinical settings due to a number of practical constraints. Consequently, HARDI data typically suffer from relatively poor spatial resolution and other effects of undersampling, thus undermining the adequacy of ensuing data analysis by means of fiber tractography.

In addition to the use of parallel imaging \cite{Feinberg2013}, the problem of long acquisition times in dMRI has been addressed using a range of post-processing solutions. In particular, most works in this direction has focused on the development of post-processing methods capable of reconstructing the dMRI signals from their partial measurements in either the $k$-space \cite{lustig2007sparse} or $q$-space. In the latter case, accelerated imaging is achieved through the use of a deliberately smaller number of diffusion encodings then necessary, while compensating for the effects of undersampling by means of properly regularized inverse solvers\index{inverse solvers}, including the recent use of deep neural networks\index{deep neural networks}: \cite{OrthogonalSlice} reconstructed the DWI from undersampled $k$-space using graph CNNs,  \cite{qspace@2015} suggested to learn diffusion metrics directly from undersampled $q$-space. In virtually all such cases, however, the directions of diffusion encoding gradients\index{directions of diffusion encoding gradients} is assumed to be given and fixed, being typically defined by means of the electrostatic repulsion (Thomas) algorithm \cite{jones1999optimal} or tessellation of platonic solids \cite{icosahedron1966}. 

Recent studies in computational imaging for inverse problems have uncovered significant benefits of simultaneously learning the forward and inverse operators, implying a concurrent estimation of the inverse (reconstruction) model along with the parameters of the acquisition system in use \cite{ours2019learningBF,haim2018depth,kellman2019datadriven}. Lately, these ideas have been used to accelerate structural MRI imaging through the use of convolutional neural networks (CNNs) that are capable of learning the optimal reconstruction model {\it together with} optimal $k$-space sampling, either point- \cite{our2019fastmri,gozcu2018learning,zhang2019reducing} or trajectory-based \cite{pilot2019weiss}. Inspired by these results, the present work introduces a general estimation framework which allows one to learn the optimal directions of diffusion-encoding gradients for HARDI along with an optimal reconstruction model required for signal reconstruction from under-sampled data.

\subsection{Main Contributions}
In this paper, we demonstrate that:
\begin{itemize}
\item using learned directions of diffusion encoding along with a learned reconstruction model leads to substantial improvements in the quality of estimated dMRI signals;
\item the proposed solution leads to improved performance of fiber tractography as an important example of dMRI-related end-tasks;
\item the learned directions of diffusion-encoding gradients generalize to dMRI datasets other than the data used for learning the directions. 
\end{itemize}

\section{Method}
\begin{figure}[!t]
    \centering
    \includegraphics[width=\linewidth]{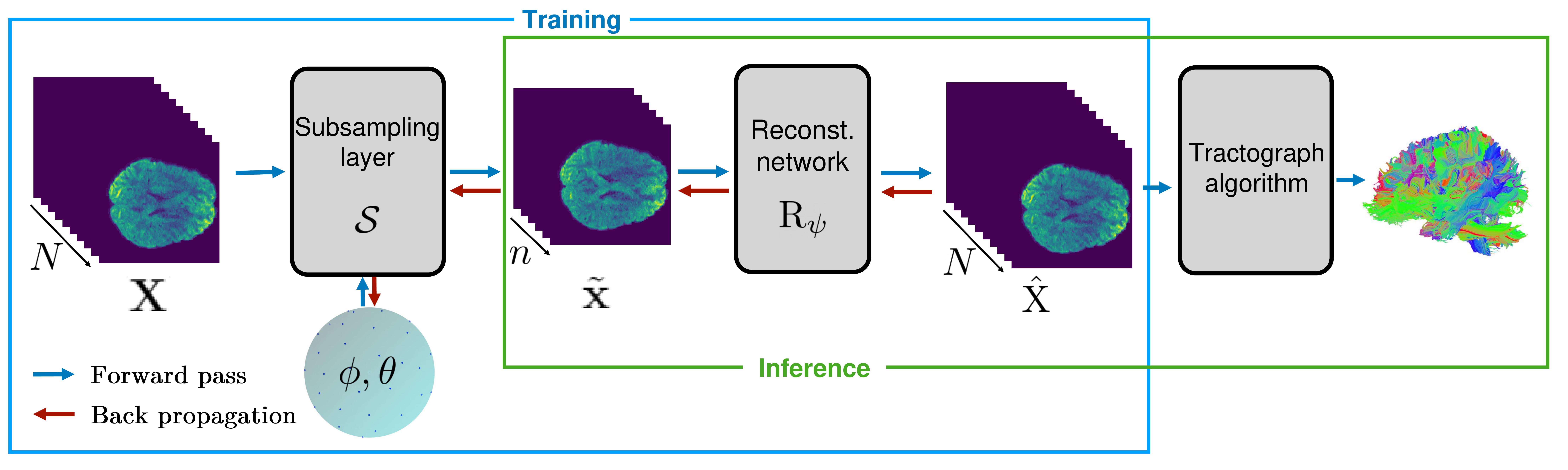}
    \caption{\small The data flow pipeline of our method. Notation is explained in the text.}
    \label{fig:model}
    \vspace{-0.5cm}
\end{figure}

The proposed method can be viewed as an end-to-end pipeline combining the forward (acquisition) and the inverse (reconstruction) models which undergo simultaneous optimization (see Fig. \ref{fig:model} for a schematic depiction). The input to the forward model, denoted as $\bb{X}$, is formed by a total of $N$ 3D diffusion-encoded volumes arranged into a 4D numerical array. The input layer is followed by a sub-sampling layer, which only uses a subset of $n \ll N$ dMRI volumes (or, equivalently, $n$ diffusion-encoding directions). The inverse model is represented by a CNN which learns the inverse mapping that goes from the measurement space (of $n$ directions) to the target space (of $N$ directions). All components of our end-to-end model are differentiable with respect to the directions of diffusion encoding (represented by their spherical coordinates $\theta$ and $\phi$), which makes the latter trainable with respect to the performance of desired end-task (e.g. fully-sampled signal reconstruction).

\subsection{Forward model: sub-sampling layer}
\label{subsec:subsampling}
The principal function of the sub-sampling layer, denoted by $\mathcal{S}_{\bb{\phi,\theta}}$, is to emulate the acquisition of dMRI measurements at a reduced number of diffusion encoding (i.e., $n$), which is referred below to as $\bb{\Tilde{X}}$. To this end, the dMRI signals used for training (and acquired at $N$ spherical points) were fit with a truncated basis of spherical harmonics. In this case, it is convenient to describe the directions of diffusion encoding in terms of their azimuth $0 \leq \phi \leq 2\pi$ and elevation $0 \leq \theta \leq \pi$, which can both be collected into vectors of length $n$. In what follows, we refer to the ratio $N/n$ as the \emph{acceleration factor} (AF). 

\subsection{Reconstruction model}
The goal of the reconstruction model is to extract the latent image ${\bb{X}}$ from the limited set of diffusion directions $\bb{\Tilde{X}}$. The resulting approximation is henceforth denoted as $\hat{\bb{X}} = R_{\bb{\psi}}(\bb{\Tilde{X}})$, where $R$ is the reconstruction model and $\bb{\psi}$ represent its learnable parameters.
Although in our implementation we chose the off-the-shelf U-NET as the reconstruction model architecture, architectural search of the optimal reconstruction model is not within the scope of this work. Our proposed algorithm can be used with any differentiable reconstruction model.

\subsection{Optimization}
\label{subsec:Optimization}
Training of the proposed pipeline is performed by simultaneously learning the diffusion gradient directions ($\phi,\theta$) and the parameters of the reconstruction model $\psi$. The training is carried out by minimizing the discrepancy between the model output image $\hat{\bb{X}}$ and the ground-truth image $\bb{X}$, meaning, by solving the optimization problem 
\begin{equation}
\min_{\bb{\psi}, \bb{\phi},\bb{\theta}} \sum_{\bb{X}} \|R_{\bb{\psi}}(\mathcal{S}_{\bb{\phi,\theta}}(\bb{X}))- \bb{X}\|_2
\label{eq:min}
\end{equation}
where the loss is summed over a training set comprising of 4D diffusion volumes of the full set of diffusion directions $\bb{X}$.

\section{Experimental evaluation}

\subsection{Dataset}
The experiments reported in this paper have been obtained using the dMRI data provided by the Human Connectome Project (HCP) \cite{HCP@2012}. The HCP database contains $1065$ brain MRI volumes acquired with $288$ diffusion-encoding directions including $90$ directions for each of the $b$ values 1000, 2000, 3000 s/mm$^2$ and the rest 18 volumes with b value of 0. For the sake of simplicity, we choose to use only the diffusion directions pertaining to the spherical shell defined by $b=1000$ s/mm$^2$. We used 868 volumes (102,000 slices) for training and 100 volumes (12,000 slices) for validation. To maintain consistency across the dataset that was originally acquired with different gradient directions schemes, we first resample the DWI data into $N=90$ predefined directions evenly distributed on the unit hemisphere using spherical harmonics.
\vspace{-0.2cm}

\subsection{Training settings}
The sub-sampling layer and the reconstruction network were trained with the Adam optimizer \cite{adam2014}. The learning rate was set to $0.001$ for the reconstruction model, while the sub-sampling layer was trained with a learning rate of $0.0001$. For the reconstruction model, we used a multi-resolution encoder-decoder network with symmetric skip connections, also known as the U-Net architecture \cite{ronneberger2015unet}. U-Net is widely-used in medical imaging tasks with many application in structural MRI reconstruction \cite{zbontar2018fastmri} and segmentation \cite{MRIsegment}. The reconstruction network has been trained in a slice-wise manner, with the image slices being concatenated back to 3D volume during the stage of inference. The input to the model is given by a slice where each of the diffusion directions $n$ is a channel and the output is the slice with $N=90$ channels. We emphasize that the scope of this work is not directed toward building the best reconstruction method, but rather demonstrating the benefit of simultaneous optimization of the acquisition-reconstruction pipeline. We perform experiments for the following acceleration factors (AF): $3, 5, 10, 15, \, \& \, 30$ that correspond to acquisitions performed with $n=30, 18, 9, 6, 3$ diffusion directions, respectively.
\vspace{-0.2cm}
\subsection{Results and discussion}

\begin{table}[!t]
    \centering
\addtolength{\tabcolsep}{1pt}
\begin{tabular}{|c|c|c|c|c|c|c|}
\hline
AF/$n$&3/30&5/18&10/9&15/6&30/3\\\hline
\hline
fixed dirs.&48.99$\pm$1.62&48.73$\pm$1.51&45.13$\pm$1.67&42.04$\pm$1.52&39.67$\pm$1.45\\
learned dirs.&\textbf{49.18$\pm$1.52}&\textbf{48.94$\pm$1.52}&\textbf{45.23$\pm$1.49}&\textbf{42.20$\pm$1.42}&\textbf{39.70$\pm$1.44}\\
\hline
\end{tabular}
\caption{\small \textbf{Quantitative results in the signal space.} The presented baselines are networks trained with \textit{fixed} and \textit{learned} directions across different acceleration factors (AF). Presented are the PSNR between the reconstructed and groundtruth DWI volumes. $n=$ number of diffusion directions.}\label{PSNR}
\vspace{-0.8cm}
\end{table}

\paragraph{Baselines.} We consider the following baselines: \textit{learned directions} (joint optimization of diffusion directions and the reconstruction network, $\bb{\phi}$ and $\bb{\theta}$ were initialized randomly), and \textit{fixed directions} (optimizing the reconstruction network alone). For the fixed regime, we use the electrostatic repulsion algorithm \cite{jones1999optimal} to design the diffusion gradient and only train the reconstruction network. In order to demonstrate the robustness of the learned diffusion directions and their practical applicability in the absence of reconstruction network, we present two additional baselines, namely, \textit{fixed w/o reconstruction} and \textit{learned w/o reconstruction}. Effectively, this implies rendering the operator $R_{\bb{\psi}}$ to be identity.
\paragraph{Metrics in signal space.} First, we evaluate the algorithm performance by comparing directly $\hat{\bb{X}}$ the ground-truth $\bb{X}$ using the PSNR (peak signal-to-noise ratio). In Table \ref{PSNR}, we present the results of the fixed \& learned directions baselines across several acceleration factors. As expected, we notice that the PSNR of the reconstructed images degrade as the acceleration factor increases. Interestingly, we can notice improvement in the case of learned directions over the fixed one demonstrating the merit of joint optimization of diffusion directions and reconstruction network.
\paragraph{Metrics in tractography space.} The performance of the proposed method has been also tested in application to fiber tractography. To this end, we applied the same tractography reconstruction to both $\hat{\bb{X}}$ \& $\bb{X}$ (concatenate with 6 originals volumes of $b=0$) and compare their tractograms. To perform tractography, we first fit the DWI to constant solid angle ODF model \cite{kamath2012generalized} and then invoke fiber tracking using EuDX \cite{garyfallidis2013EuDX} using the implementation available in the \texttt{dipy} package.
To compare the tractograms, we used the Bhattacharyya distance \cite{Bhattacharyya1946} over \textit{bundles} (a collection of individual fibres), we compute it over $15$ different bundles and compute the average distance. Table \ref{BD-table} presents a comparison of the results obtained from fixed \& learned directions baselines. We evaluate the quality of the learned directions both with and without using the reconstruction network across few acceleration factors. Visual results of the entire tractogram can be seen in Fig. \ref{fig:tract} and for specific bundles in Fig. \ref{fig:bundle} (visualization for all acceleration factor are presented in the supplementary material, see Figs. \ref{fig:supp-tract-side}, \ref{fig:supp-tract-up}, \ref{fig:supp-bundle-MCP} \& \ref{fig:supp-bundle-CS_L}). Based on the obtained quantitative and qualitative results, the following are the observations 
\begin{itemize}
    \item The use of supervised learning for reconstruction gave significant improvement of $0.186-1.346$ points in Bhattacharyya distance and a clear noticeable improvement in the qualitative results presented in  Figs. \ref{fig:supp-tract-side}, \ref{fig:supp-tract-up}, \ref{fig:supp-bundle-MCP} \& \ref{fig:supp-bundle-CS_L}.
    \item Joint optimization of the diffusion directions and the reconstruction gave a further sizeable improvement of $0.015-0.478$ points in Bhattacharyya distance.
    \item A particularly appealing result is that, we notice even in the absence of reconstruction network, the learned directions yield an improvement of $0.076-0.735$ points in Bhattacharyya distance when compared to the fixed directions. A possible reason for this improvement is because joint optimization of the reconstruction network together with the diffusion directions enforces metric that is learned from the data in the direction space. This metric is more meaningful than minimizing the electrostatic energy. Our learned diffusion directions can be simply translated into diffusion gradients and readily deployed onto real MRI scanners to achieve sizeable acceleration.
\end{itemize}

\begin{table}[!t]
    \centering
\addtolength{\tabcolsep}{1pt}

\begin{tabular}{|c|c|c|c|c|c|c|c|c|c|c|}
\hline
&\multicolumn{5}{|c|}{Bhattacharyya distance}\\
\hline
AF/$n$&3/30&5/18&10/9&15/6&30/3\\\hline
\hline
fixed w/o reconst.&0.329$\pm$0.032&0.539$\pm$0.057&0.685$\pm$0.058&1.802$\pm$0.190&--\\
learned w/o reconst.&0.253$\pm$0.024&0.327$\pm$0.032&0.514$\pm$0.055&1.067$\pm$0.092&--\\
\hline 
fixed w/ reconst.&0.143$\pm$0.029&0.172$\pm$0.032&0.264$\pm$0.031&0.456$\pm$0.048&0.860$\pm$0.097\\
learned w/ reconst.&\textbf{0.128$\pm$0.022}&\textbf{0.151$\pm$0.027}&\textbf{0.205$\pm$0.029}&\textbf{0.335$\pm$0.035}&\textbf{0.382$\pm$0.036}\\
\hline
\end{tabular}
\caption{\small \textbf{Quantitative results in the tractography space.} The presented baselines are networks trained with \textit{fixed} and \textit{learned} directions across different acceleration factors (AF). Presented below are the Bhattacharyya distance (BD) between the tractographies obtained from the groundtruth and reconstructed DWI volumes. The distance is averaged over $15$ bundles. $n=$ number of diffusion directions.}\label{BD-table}
\vspace{-1cm}
\end{table}

\begin{figure}[!hb]
   \vspace{-0.5cm}
   \centering
   \addtolength{\tabcolsep}{-5.5pt}
\begin{tabular}{c c c c c}
fixed dir.& learned dir. & fixed dir. & learned dir. & \multirow{2}{*}{groundtruth}\\
w/o reconst. & w/o reconst. & w/ reconst. & w/ reconst. &  \\
\includegraphics[width=0.22\textwidth]{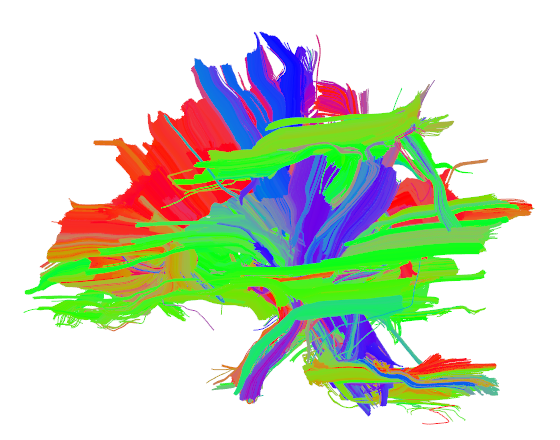}&
\includegraphics[width=0.22\textwidth]{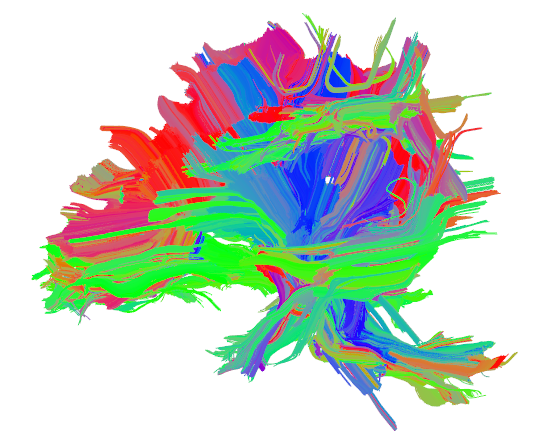}&
\includegraphics[width=0.22\textwidth]{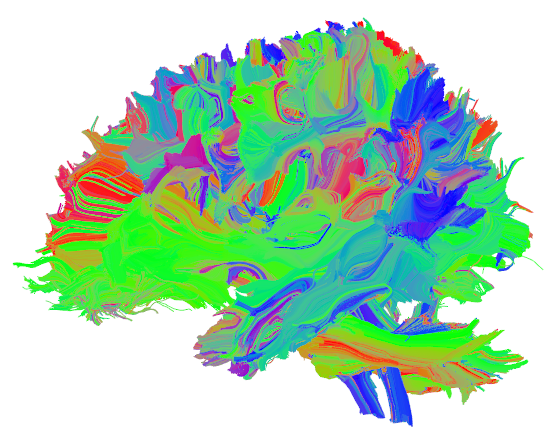}&
\includegraphics[width=0.22\textwidth]{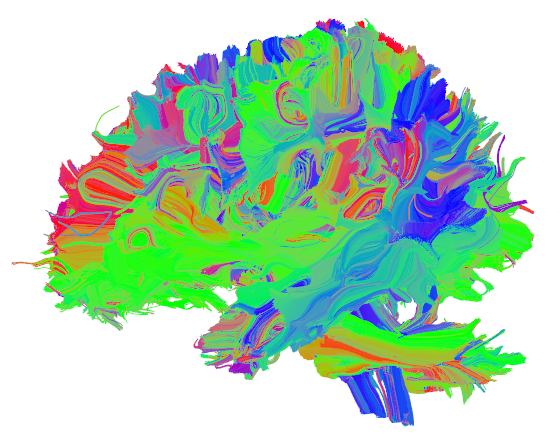}&
\includegraphics[width=0.22\textwidth]{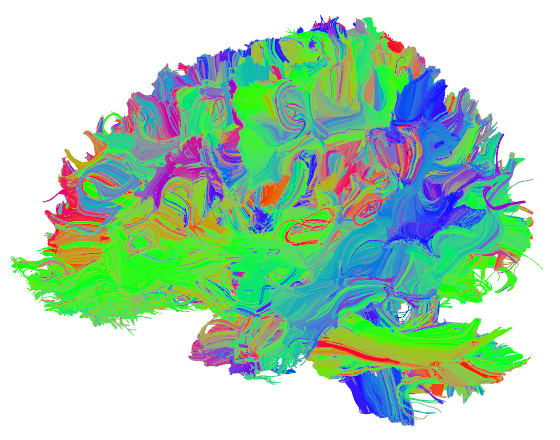}\\
\includegraphics[angle=270,width=0.2\textwidth]{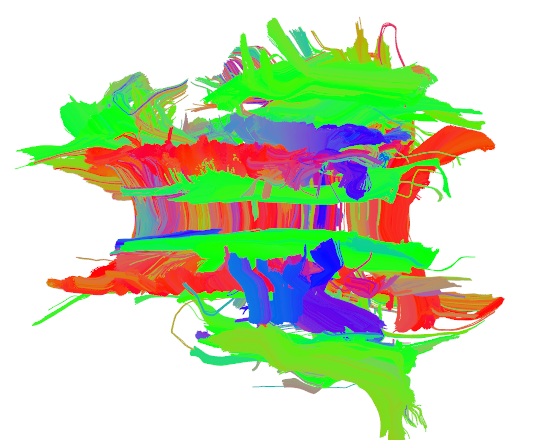}&
\includegraphics[angle=270,width=0.2\textwidth]{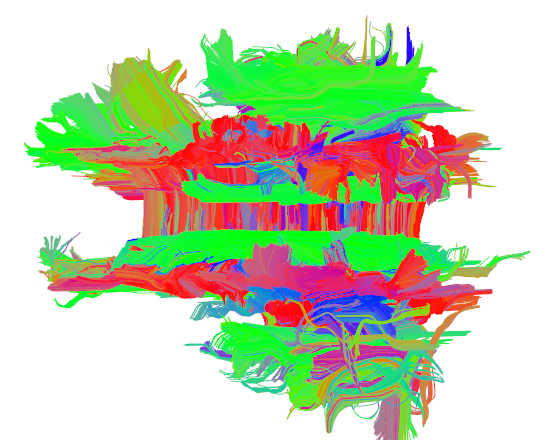}&
\includegraphics[angle=270,width=0.2\textwidth]{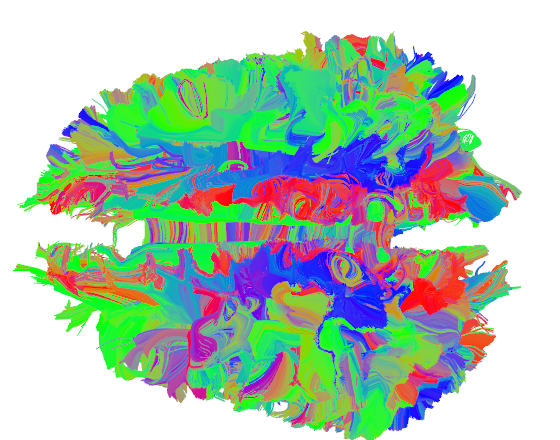}&
\includegraphics[angle=270,width=0.2\textwidth]{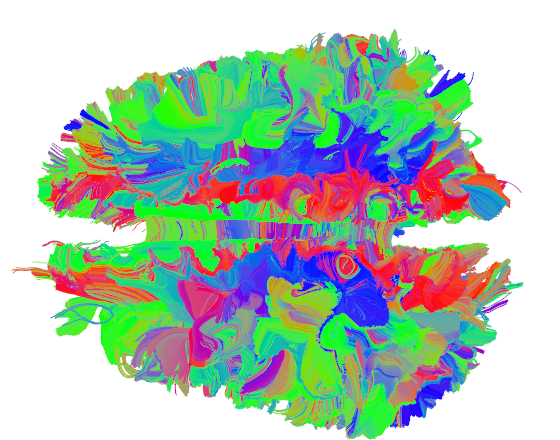}&
\includegraphics[angle=270,width=0.2\textwidth]{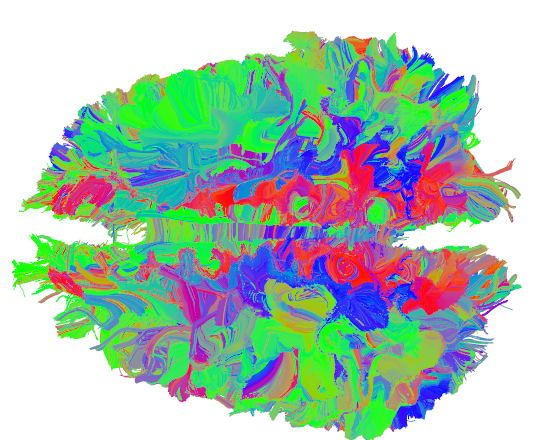}\\

\end{tabular}
    \caption{\textbf{Qualitative evaluation of the tractogram of a test sample from the human connectome project.} Presented below are the tractograms achieved using $6$ gradient directions (AF=15) for fixed/learned directions with/without the reconstruction network, and the corresponding groundtruth tractogram. Digital zoom-in is recommended. Top: side-view, bottom: top-view.}
    \label{fig:tract} 
\vspace{-0.5cm}
\end{figure}

\begin{figure}[!b]
   \centering
   \addtolength{\tabcolsep}{-5.5pt}
\begin{tabular}{c c c c c}
\multirow{2}{*}{Bundle}& fixed dir. & learned dir. & fixed dir. & learned dir.\\
& w/o reconst. & w/o reconst. & w/ reconst. & w/ reconst.\\
\rotatebox{90}{\hspace{1.3cm} CS-L}&
\includegraphics[width=0.24\textwidth]{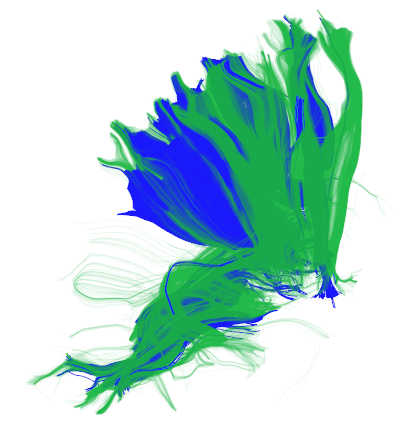}&
\includegraphics[width=0.24\textwidth]{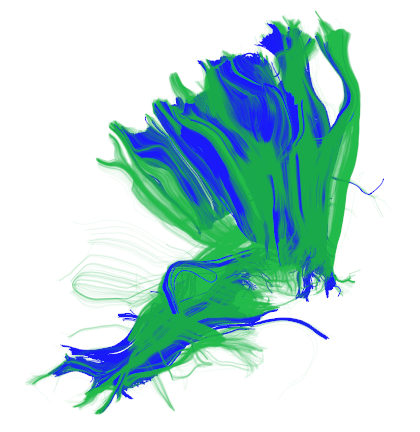}&
\includegraphics[width=0.24\textwidth]{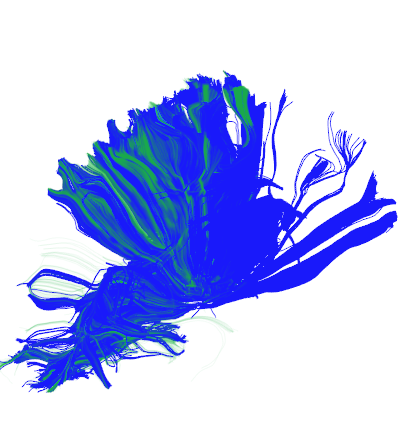}&
\includegraphics[width=0.24\textwidth]{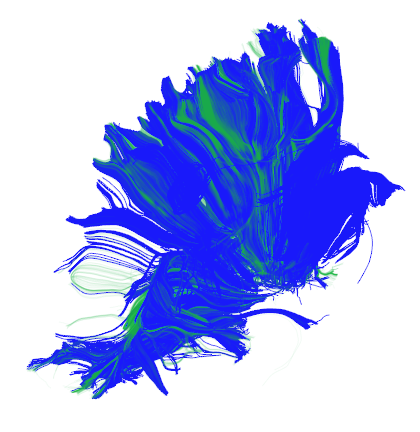}\\
\rotatebox{90}{\hspace{1.3cm} MCP}&
\includegraphics[width=0.24\textwidth]{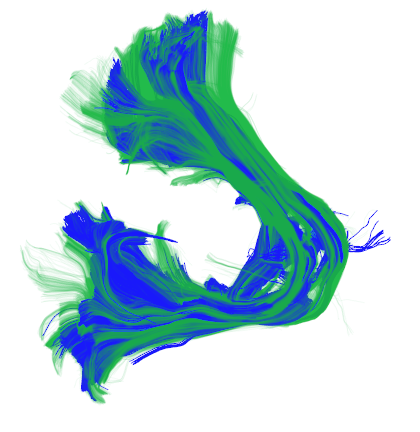}&
\includegraphics[width=0.24\textwidth]{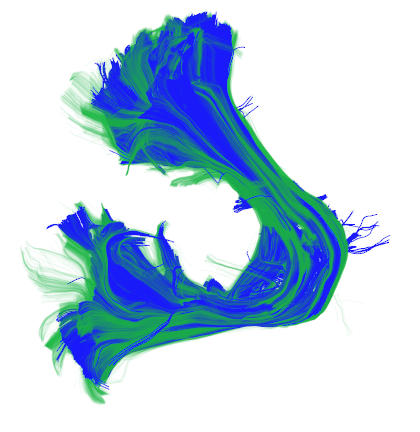}&
\includegraphics[width=0.24\textwidth]{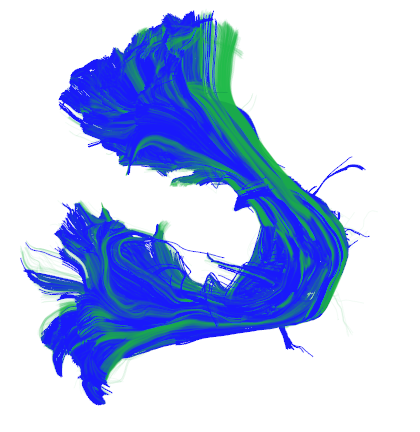}&
\includegraphics[width=0.24\textwidth]{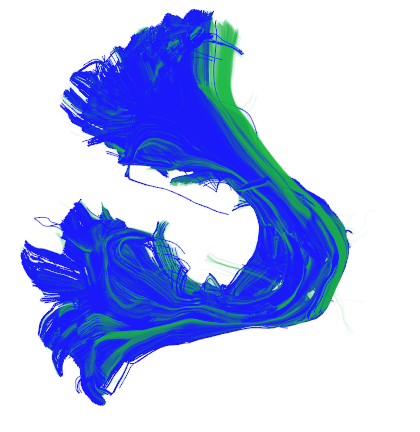}\\
\end{tabular}
    \caption{\textbf{Visualization of CS-L \& middle cerebellar peduncle bundles} of a test volume in the HCP dataset after tractography. Results depicted are obtained using fixed/learned diffusion directions (AF=3, $n=30$), both with/without reconstruction. Colored in blue and green are the reconstructed and groundtruth bundles respectively.}
    \label{fig:bundle} 
\vspace{-0.5cm}
\end{figure}

\vspace{-0.4cm}
\subsubsection{Learned directions.}
\begin{wrapfigure}{r}{0.50\textwidth}
\vspace{-8.5pt}
\addtolength{\tabcolsep}{-8pt}
\includegraphics[width=0.24\textwidth]{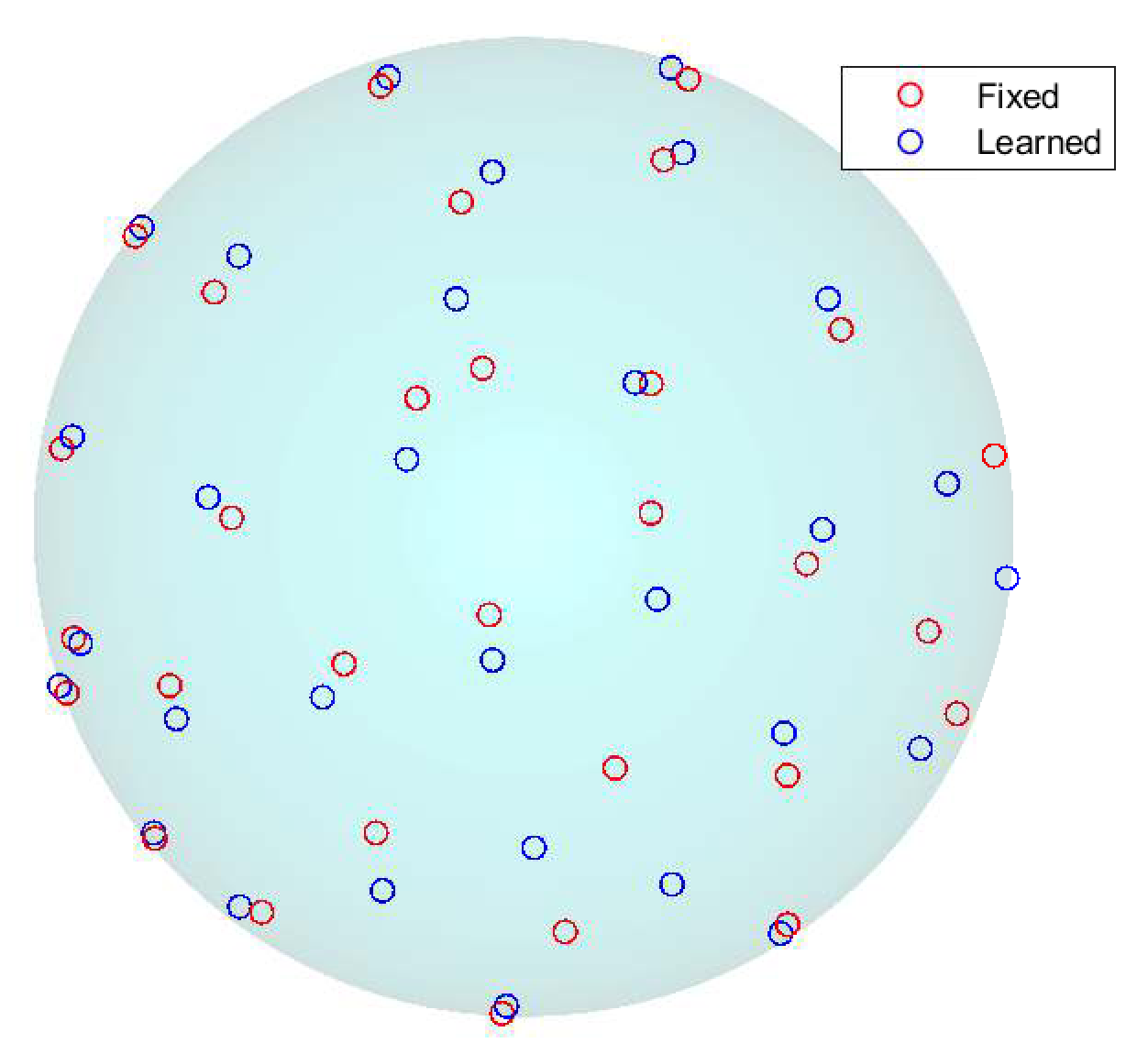}
\includegraphics[width=0.24\textwidth]{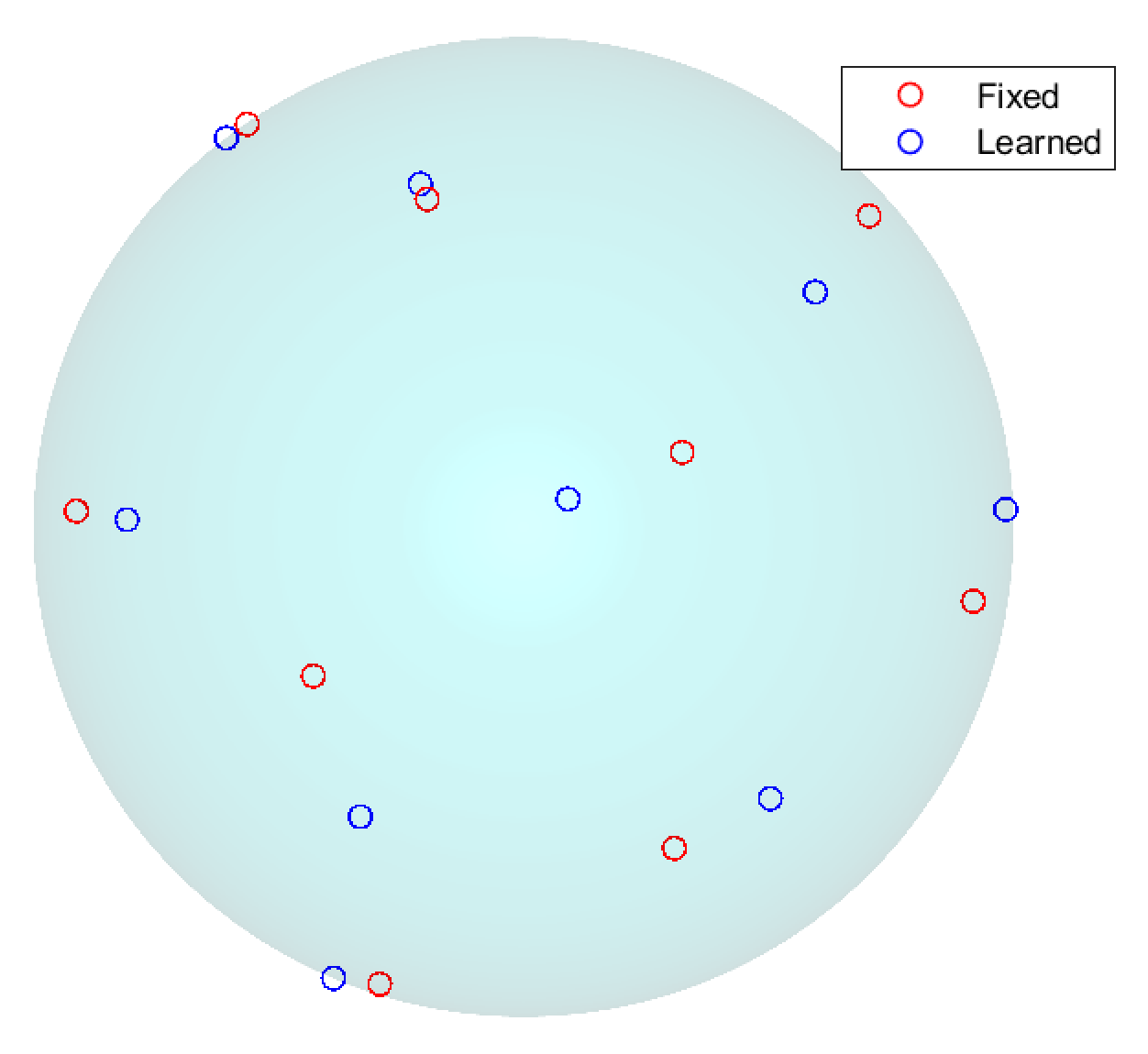}
\vspace{-14pt}
\label{fig:dir-plot}
\end{wrapfigure}
A comparison of the fixed (red) and learned directions (blue) plotted on a hemi-sphere is visualized in the adjacent figure. The left and right figures correspond to AF=3 ($n$=30) and AF=10 ($n$=9) respectively. We can notice that although both fixed and learned directions look similar uniformly covering the hemisphere, our results presented above results suggest that the learned directions hit the right spots that yield better reconstruction and tractography performance. This implies that the joint optimization leads to an importance sampling in the diffusion space.


\vspace{-0.5cm}
\subsubsection{Tractometer, ISMRM challenge.}
To further demonstrate the robustness, deployability and generalization capability of our learned diffusion directions, we test our learned diffusion directions on the ISMRM tractography challenge DWI phantom dataset \cite{ISMRM_data}. We evaluated the results using the \textit{tractometer tool} \cite{Tractometer}. For the evaluation, we first sub-sampled the original DWI volume using the fixed and learned diffusion directions that we obtain from the human connectome dataset. Then, we applied the tractography algorithms to the resulting sub-sampled DWI volumes and quantitative evaluate using the tractometer tool. The results depicted in Table \ref{ISMRM-table} demonstrate that the learned directions outperform the fixed directions in most of the metrics across multiple acceleration factors. Visual results of this experiments are presented in Fig. \ref{fig:supp-ISMRM} in the supplementary material.

\begin{table*}[!t]
    \center
\addtolength{\tabcolsep}{1pt}

\begin{tabular}{|c|c|c|c|c|c|c|c|c|c|}
\hline
n&&VC (+)&IC (-)&NC (-)&VB (+)&IB (-)&OL (+)&OR (-)&F1(+)\\
\hline
\multirow{2}{*}{3}&Fixed&45.32\%&54.67\%&\textbf{0.000}\%&20&\textbf{71}&15.74\%&\textbf{23.59}\%&22.88\%\\
&Learned&\textbf{52.48\%}&\textbf{47.51\%}&0.004\%&\textbf{23}&73&\textbf{23.51\%}&29.66\%&\textbf{32.46\%}\\
\hline 
\multirow{2}{*}{5}&Fixed&56.01\%&43.98\%&\textbf{0.000\%}&21&60&17.32\%&26.83\%&25.82\%\\
&Learned&\textbf{60.26\%}&\textbf{39.73\%}&\textbf{0.000\%}&\textbf{22}&\textbf{60}&\textbf{18.32\%}&\textbf{24.67\%}&\textbf{26.89\%}\\
\hline
\multirow{2}{*}{10}&Fixed.&58.86\%&41.13\%&\textbf{0.000\%}&19&43&11.55\%&18.39\%&18.81\%\\
&Learned&\textbf{61.42\%}&\textbf{38.57\%}&\textbf{0.000\%}&\textbf{21}&\textbf{42}&\textbf{12.13\%}&\textbf{17.51\%}&\textbf{19.54\%}\\
\hline
\multirow{2}{*}{15}&Fixed&26.15\%&73.84\%&\textbf{0.000}\%&\textbf{8}&20&\textbf{1.26}\%&6.22\%&\textbf{2.35\%}\\
&Learned&\textbf{28.06\%}&\textbf{71.93\%}&\textbf{0.000}\%&5&\textbf{17}&0.76\%&\textbf{3.21\%}&1.45\%\\
\hline
\end{tabular}
\caption{\small \textbf{Quantitative results on the ISMRM challenge dataset using the tractometer tool.} Depicted below are the results obtained using the fixed and learned directions \textit{without} the reconstruction network across different AFs (presented with the $n$ corresponding to each AF). VC: valid connections(+), IC: invalid connections(-), NC: non-connections(-), VB: valid bundles(+), IB: invalid bundles(-), OL: overlap(+), OR: overreach(-), F1 score(+). +/- indicate higher or lower score is better.}\label{ISMRM-table}
\vspace{-0.8cm}
\end{table*}
\vspace{-0.3cm}

\section{Conclusion}
\vspace{-0.3cm}

We demonstrated, as a proof-of-concept, that the learning-based design of diffusion gradient design in diffusion MR imaging leads to better tractography when compared to the off-the-shelf designs. To the best of our knowledge, this is the first attempt of data-driven design of diffusion gradient directions in diffusion MRI. We trained and evaluated the performance of the learned designs and reconstruction networks on the human connectome project dataset, both in signal and tractography domains. We evaluated the generalization of the learned designs on the ISMRM challenge phantom, and observed that the learned designs consistently outperform the hand-crafted ones at different acceleration factors. Therefore, we believe that the learned directions can be deployed standalone onto real machines without the reconstruction network to already improve the end-task performance (full signal reconstruction, tractography).
We defer the following important aspects to future work:
\begin{itemize}
    \item In this work, we limited our attention to only diffusion directions lying on one shell ($b=1000)$. Allowing the diffusion gradient directions on multiple shell might lead to further improvement in the end-task performance.
    \item As can be inferred from the PSNR results presented in Table \ref{PSNR} and the tractogram metrics in Tables \ref{BD-table} \& \ref{ISMRM-table}, the discrepancy in the DWI domain is not directly related to the end-task performance (in our case, tractography). This leads us to assume that optimizing directly for the end-task (or at least, closer to it) will allow for better sampling designs that are needed for the end-task and can further push the acceleration factor vs. performance trade-offs.
\end{itemize}
%
%
\bibliographystyle{splncs04}
\bibliography{refs}

\newpage
\section{Supplementary Materials}

\begin{figure*}[!h]
   \centering
\begin{tabular}{c c c c c}
n& fixed dir. no rec.& learned dir. no rec.& fixed dir. no rec.& learned dir. no rec.\\
\rotatebox{90}{\hspace{0.3cm} Grountruth}&
\multicolumn{2}{c}{\includegraphics[width=0.24\textwidth]{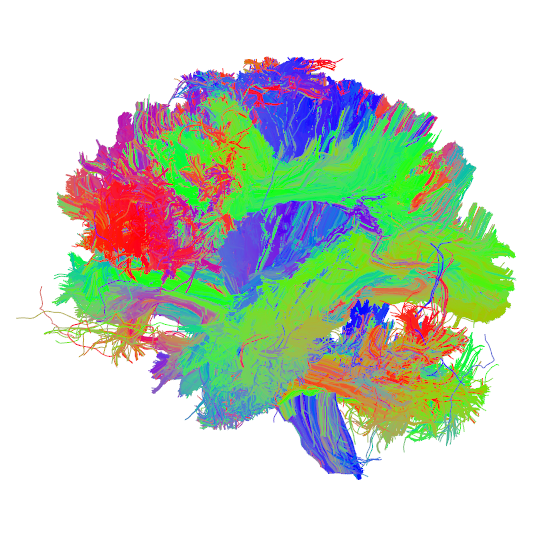}}&
\multicolumn{2}{c}{\includegraphics[angle=90,width=0.24\textwidth]{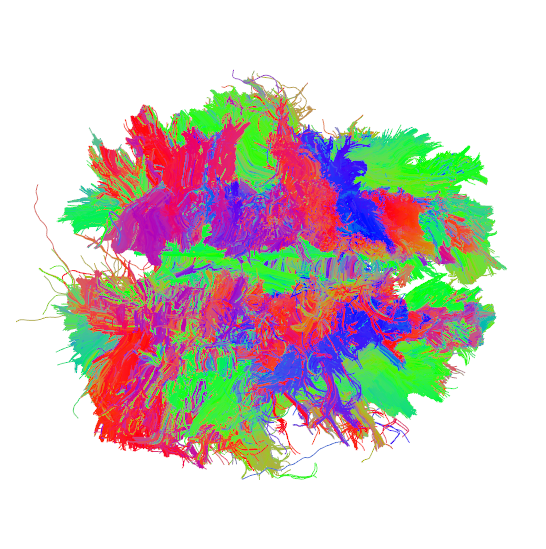}}\\
\vspace{-0.4cm}
\rotatebox{90}{\hspace{1.3cm} 30}&
\includegraphics[width=0.24\textwidth]{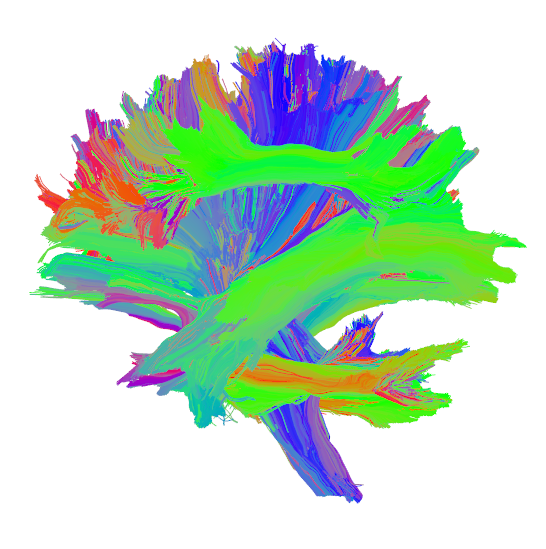}&
\includegraphics[width=0.24\textwidth]{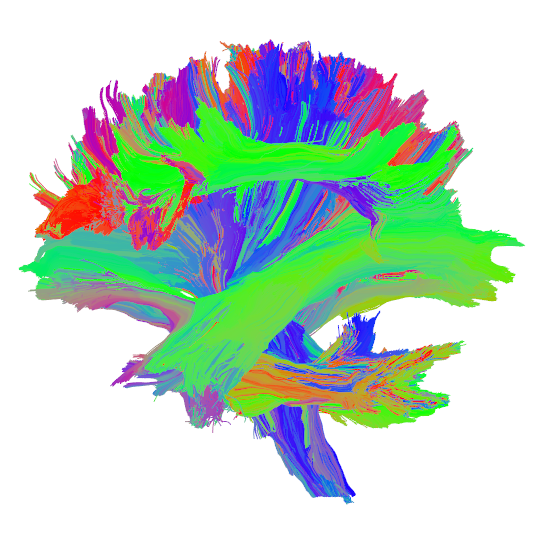}&
\includegraphics[angle=90,width=0.24\textwidth]{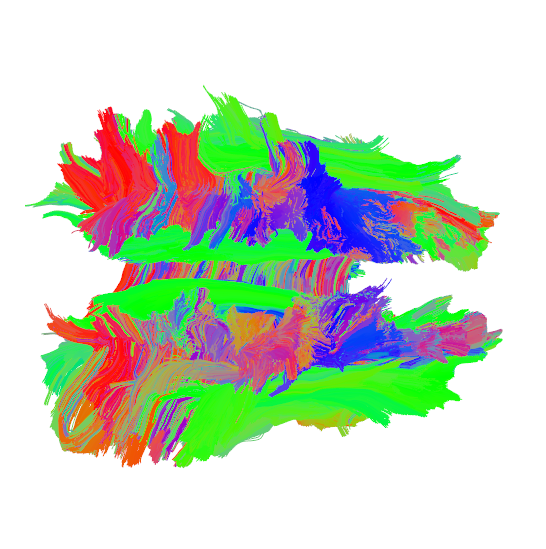}&
\includegraphics[angle=90,width=0.24\textwidth]{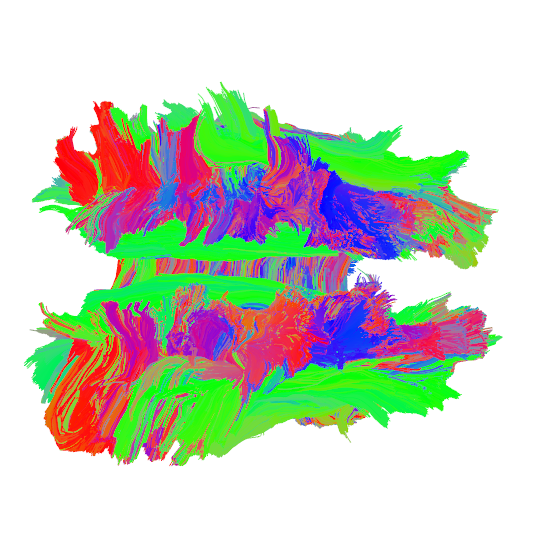}\\
\vspace{-0.4cm}
\rotatebox{90}{\hspace{1.3cm} 18}&
\includegraphics[width=0.24\textwidth]{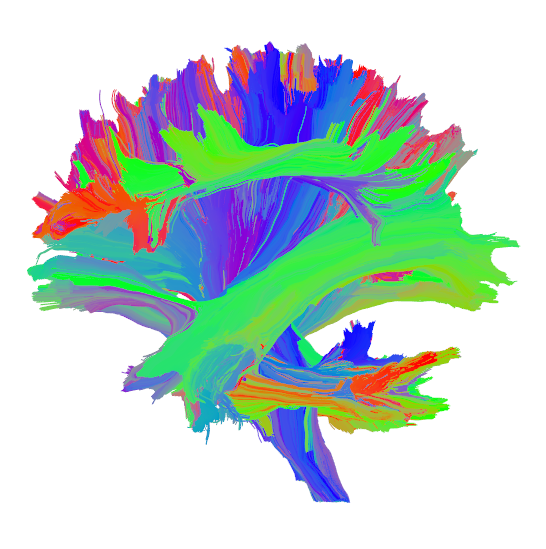}&
\includegraphics[width=0.24\textwidth]{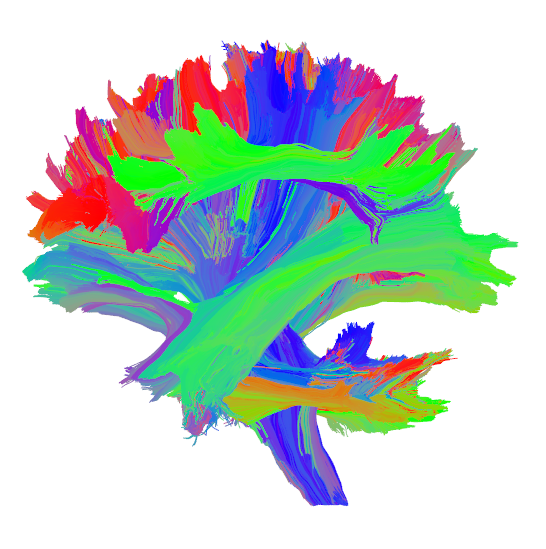}&
\includegraphics[angle=90,width=0.24\textwidth]{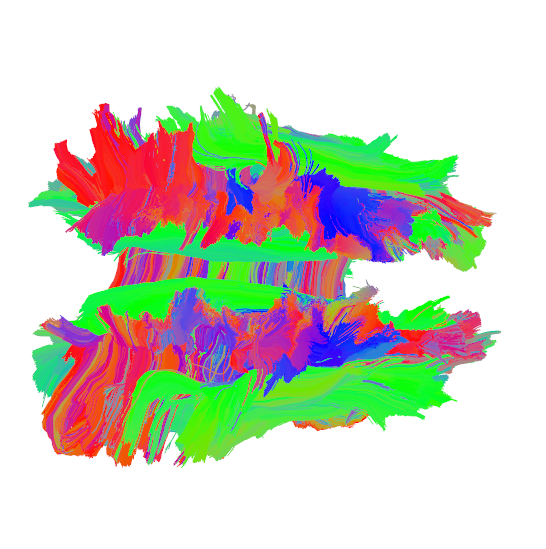}&
\includegraphics[angle=90,width=0.24\textwidth]{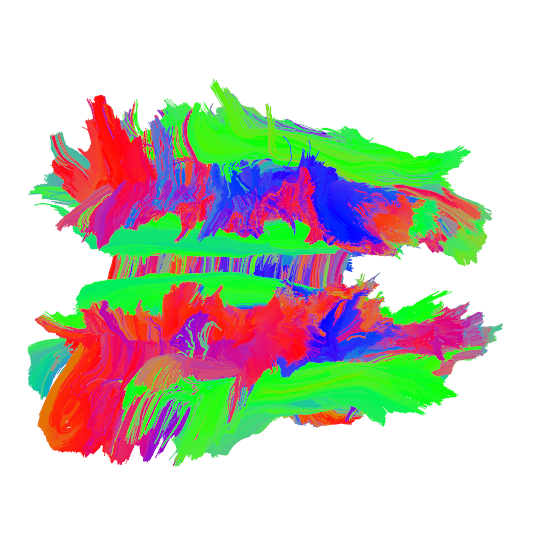}\\
\vspace{-0.4cm}
\rotatebox{90}{\hspace{1.3cm} 9}&
\includegraphics[width=0.24\textwidth]{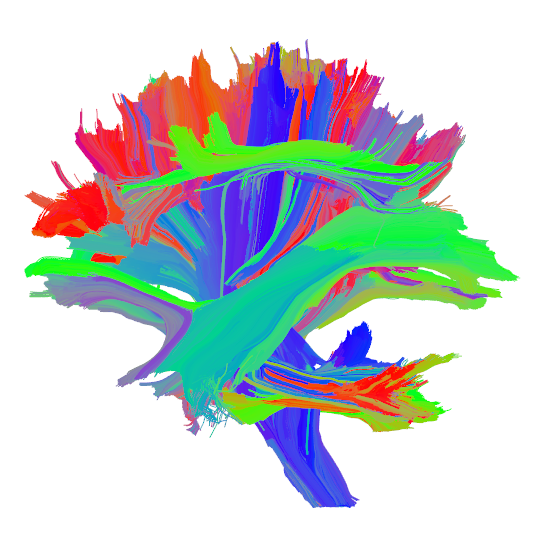}&
\includegraphics[width=0.24\textwidth]{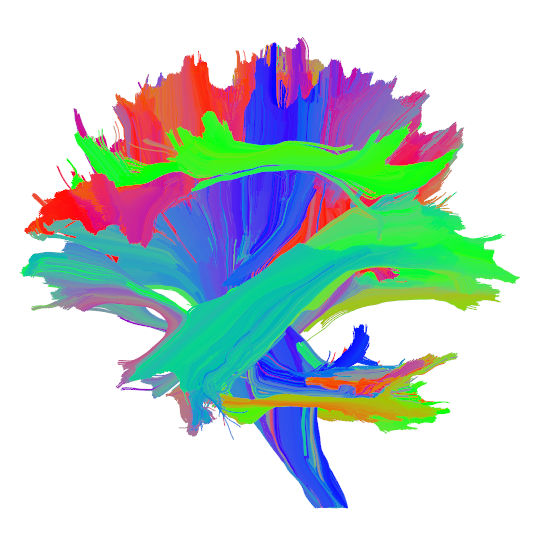}&
\includegraphics[angle=90,width=0.24\textwidth]{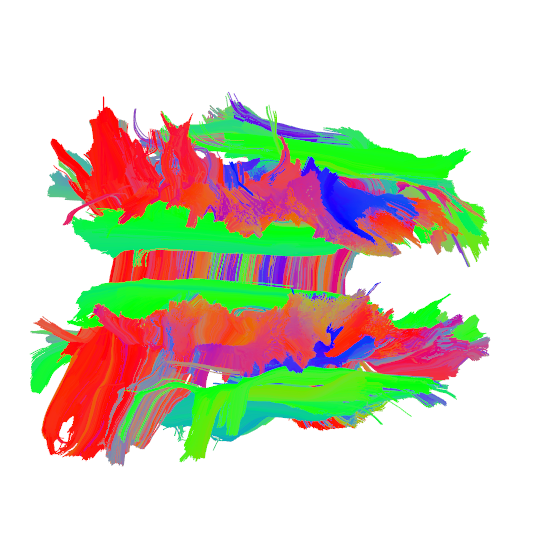}&
\includegraphics[angle=90,width=0.24\textwidth]{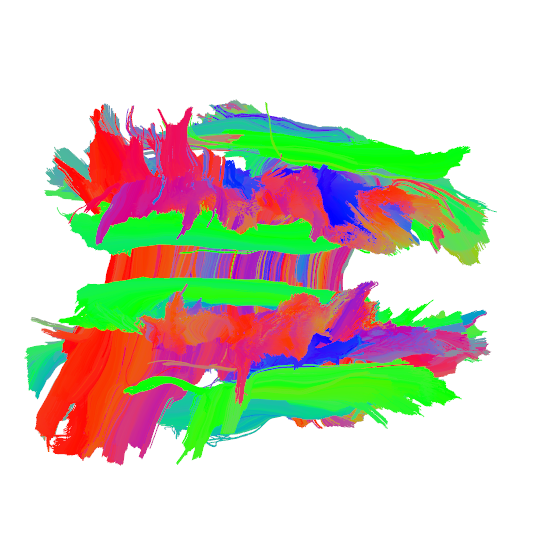}\\
\vspace{-0.4cm}
\rotatebox{90}{\hspace{1.3cm} 6}&
\includegraphics[width=0.24\textwidth]{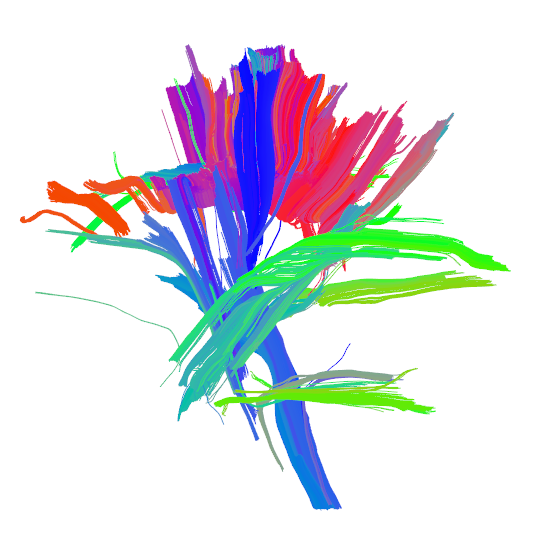}&
\includegraphics[width=0.24\textwidth]{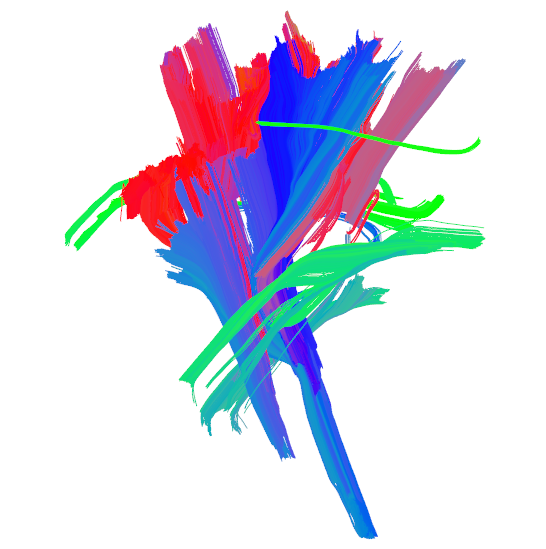}&
\includegraphics[angle=90,width=0.24\textwidth]{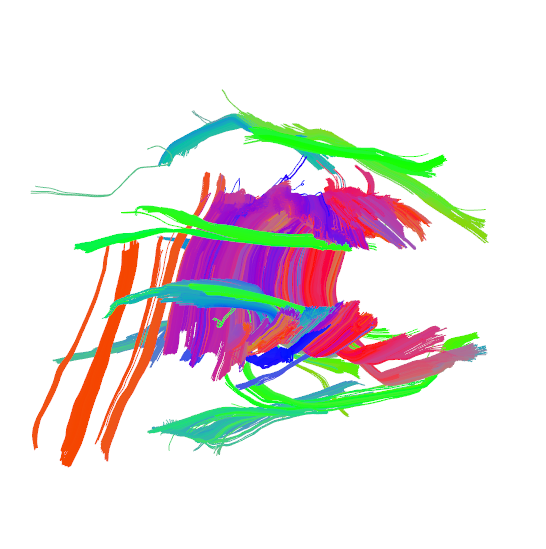}&
\includegraphics[angle=90,width=0.24\textwidth]{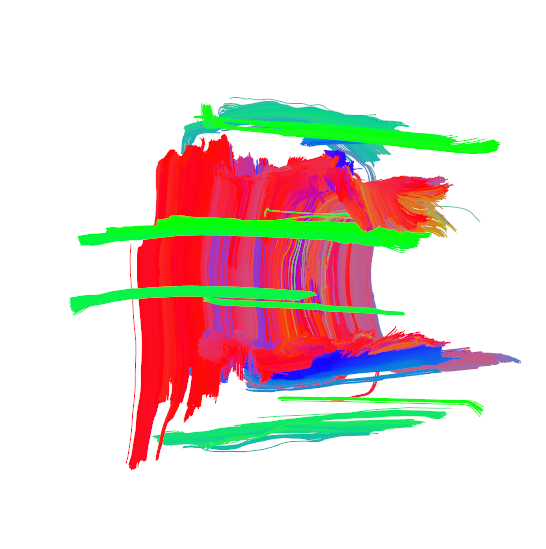}\\
\vspace{-0.4cm}
\end{tabular}
    \caption{\textbf{ISMRM fiber cup tractogram.} Visual comparison of the learned and fixed diffusion parameters applied on diffusion data from the ISMRM fiber cup challenge. Note that we only use the learned directions and not the reconstruction network. Presented in figures are the side \& top views, without reconstruction. Left column depicts the number of diffusion directions used. The first two columns and last two columns depict the side-view and top-view of the same tractogram. Note that the above depicted figures are obtained using just the learned diffusion sampling without the reconstruction network.}
    \label{fig:supp-ISMRM} 
\end{figure*}

\begin{figure*}
   \centering
   \addtolength{\tabcolsep}{-5.5pt}
\begin{tabular}{c c c c c}
AF (n)&Fixed-no rec.&Learned-no rec.&Fixed-with rec&Learned-with rec.\\
\rotatebox{90}{\hspace{0.1cm} Grountruth}&
\multicolumn{4}{c}{\includegraphics[width=0.24\textwidth]{viz/gt_side.png}}\\
\rotatebox{90}{\hspace{0.8cm} 3 ($n=30$)}&
\includegraphics[width=0.24\textwidth]{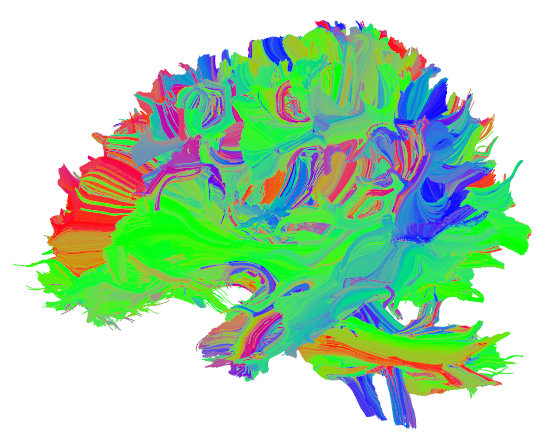}&
\includegraphics[width=0.24\textwidth]{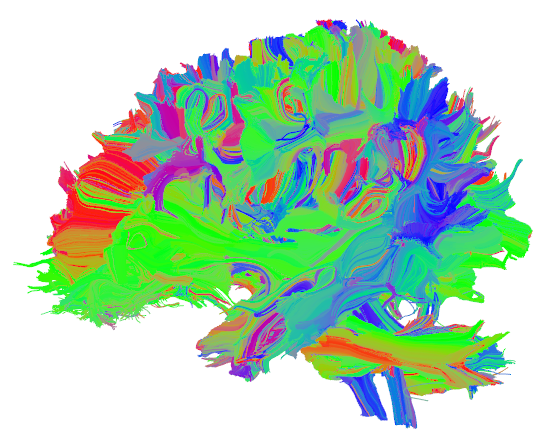}&
\includegraphics[width=0.24\textwidth]{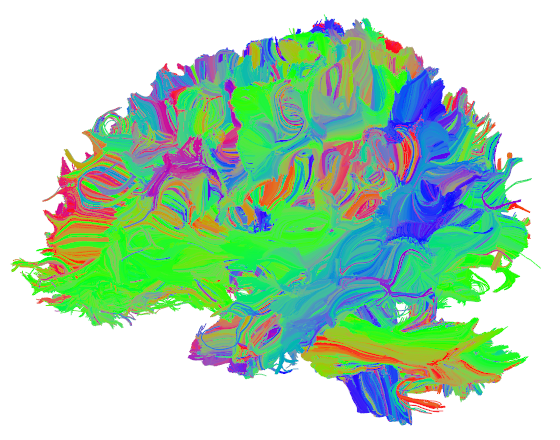}&
\includegraphics[width=0.24\textwidth]{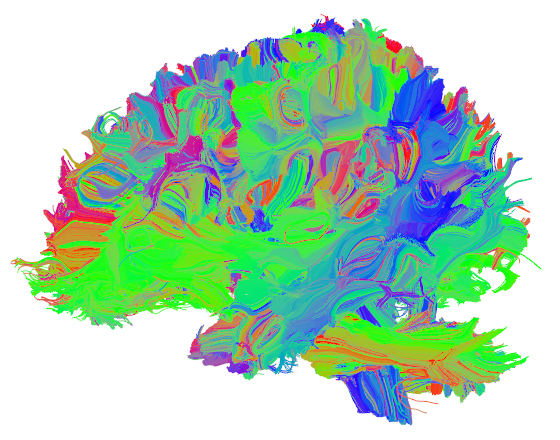}\\
\rotatebox{90}{\hspace{0.8cm} 5 ($n=18$)}&
\includegraphics[width=0.24\textwidth]{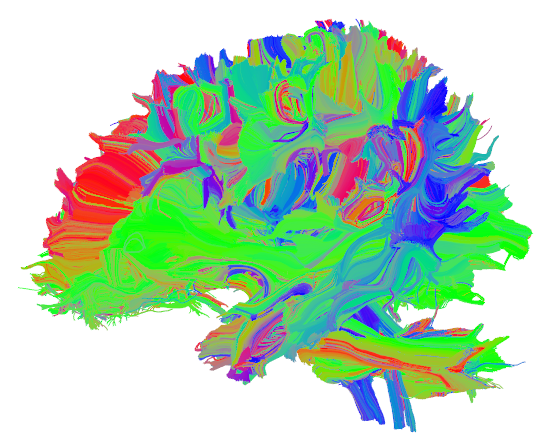}&
\includegraphics[width=0.24\textwidth]{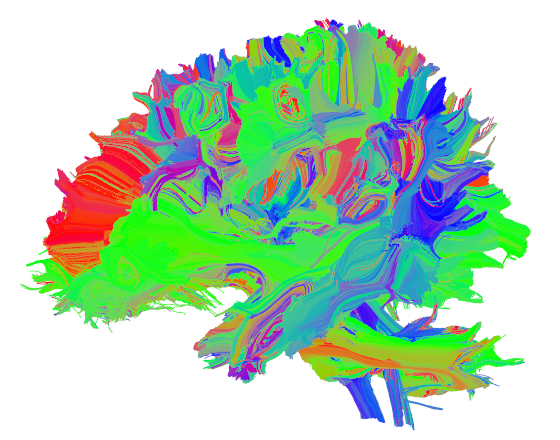}&
\includegraphics[width=0.24\textwidth]{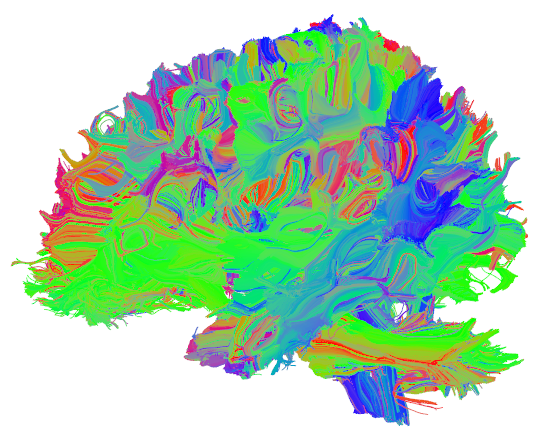}&
\includegraphics[width=0.24\textwidth]{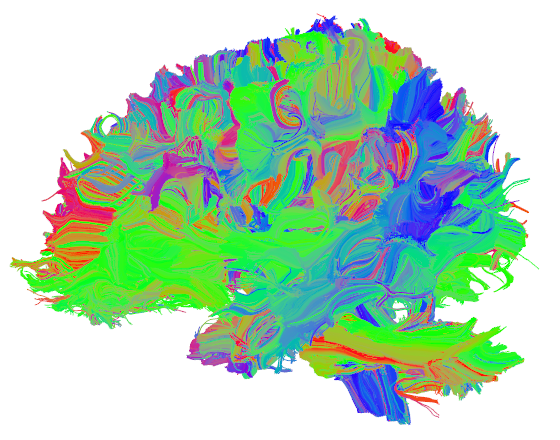}\\
\rotatebox{90}{\hspace{0.8cm} 10 ($n=9$)}&
\includegraphics[width=0.24\textwidth]{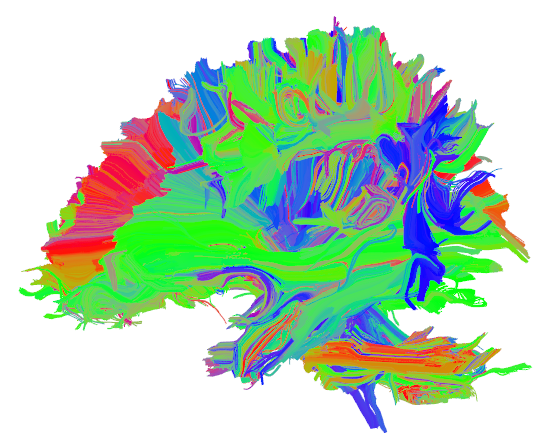}&
\includegraphics[width=0.24\textwidth]{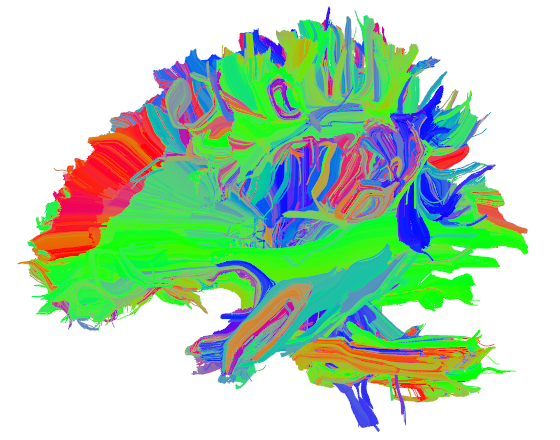}&
\includegraphics[width=0.24\textwidth]{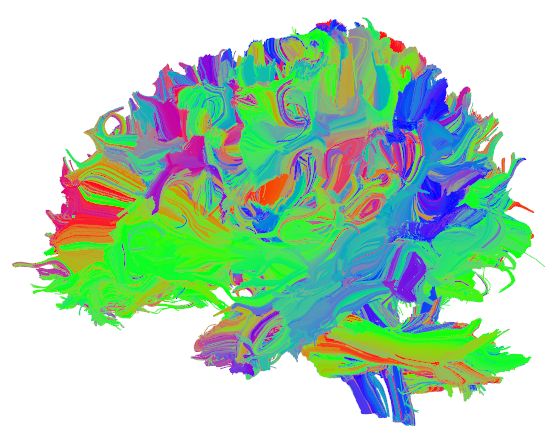}&
\includegraphics[width=0.24\textwidth]{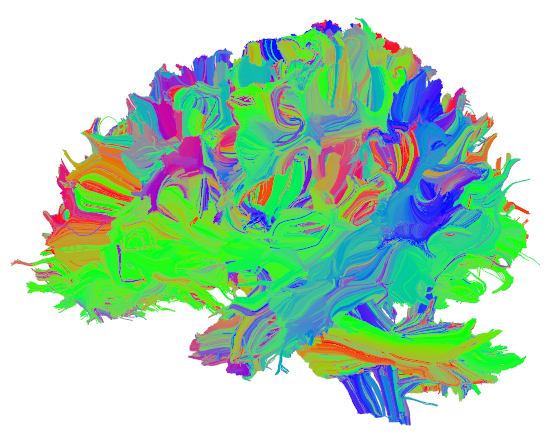}\\
\rotatebox{90}{\hspace{0.8cm} 15 ($n=6$)}&
\includegraphics[width=0.24\textwidth]{viz/15/fixed_sub_side.png}&
\includegraphics[width=0.24\textwidth]{viz/15/learned_sub_side.png}&
\includegraphics[width=0.24\textwidth]{viz/15/fixed_side.png}&
\includegraphics[width=0.24\textwidth]{viz/15/learned_side.png}\\
\rotatebox{90}{\hspace{0.8cm} 30 ($n=3$)}&
\includegraphics[width=0.24\textwidth]{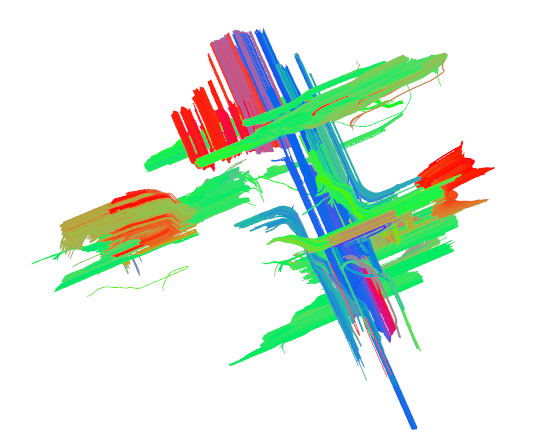}&
\includegraphics[width=0.24\textwidth]{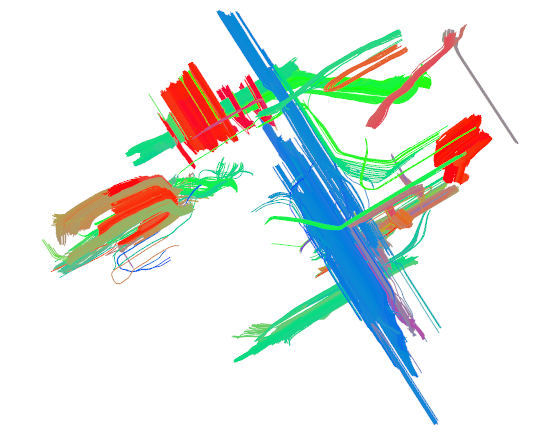}&
\includegraphics[width=0.24\textwidth]{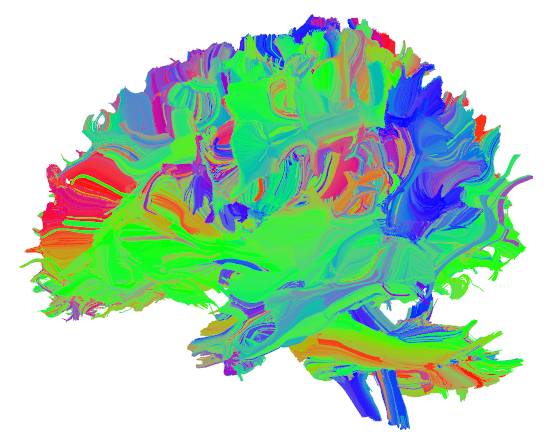}&
\includegraphics[width=0.24\textwidth]{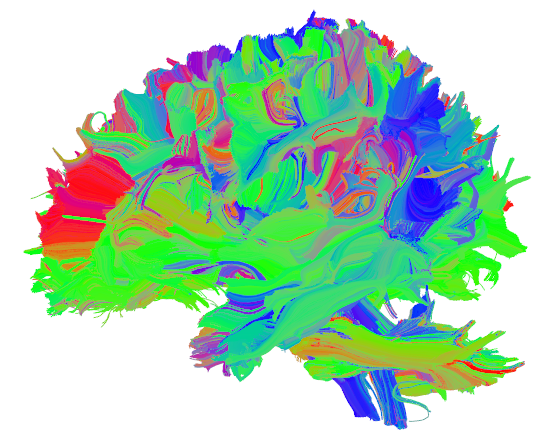}\\
\end{tabular}
    \caption{\textbf{Tractogram: side-view.} Visual comparison of the full tractogram achieved from the DWI volume using fixed \& learned directions, with \& without reconstruction. Left-most column depicts the corresponding acceleration factor and the corresponding $n$ (number of diffusion directions).}
    \label{fig:supp-tract-side} 
\end{figure*}

\begin{figure*}
   \centering
   \addtolength{\tabcolsep}{1.2pt}
\begin{tabular}{c c c c c}
AF (n)&Fixed-no rec.&Learned-no rec.&Fixed-with rec&Learned-with rec.\\
\rotatebox{90}{\hspace{-2.0cm} Groundtruth}&
\multicolumn{4}{c}{\includegraphics[width=0.24\textwidth,angle=270]{viz/gt_up.png}}\\
\rotatebox{90}{\hspace{-2.0cm} 3 ($n=30$)}&
\includegraphics[width=0.24\textwidth,angle=270]{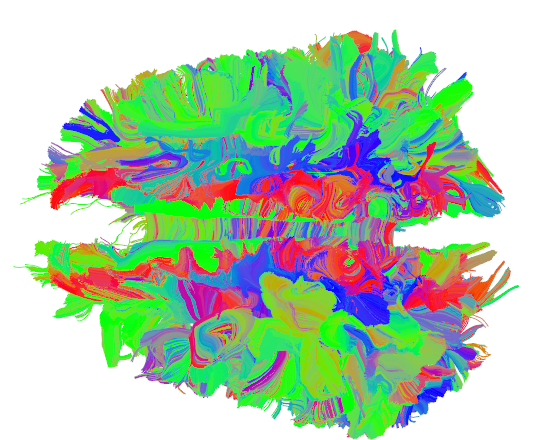}&
\includegraphics[width=0.24\textwidth,angle=270]{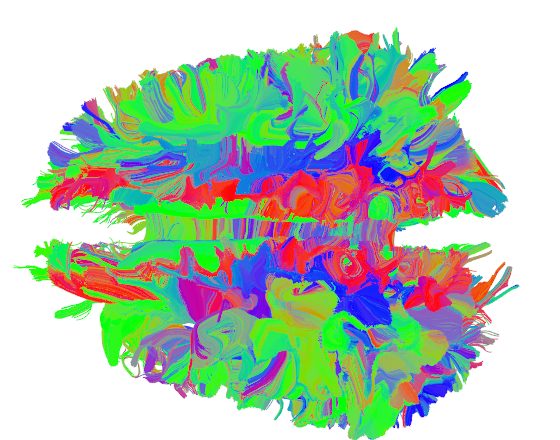}&
\includegraphics[width=0.24\textwidth,angle=270]{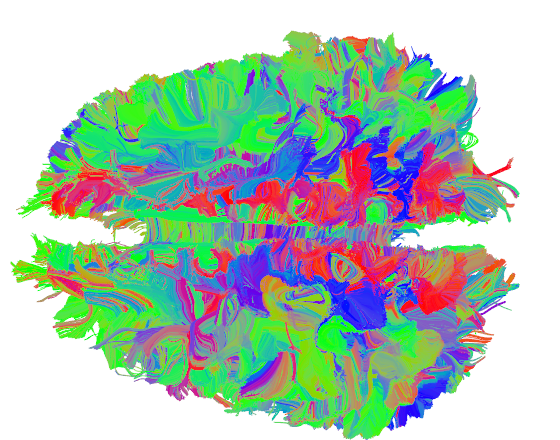}&
\includegraphics[width=0.24\textwidth,angle=270]{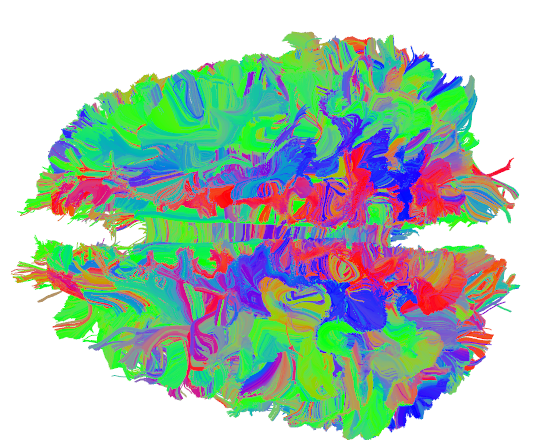}\\
\vspace{-0.24cm}
\rotatebox{90}{\hspace{-2.0cm} 5 ($n=18$)}&
\includegraphics[width=0.24\textwidth,angle=270]{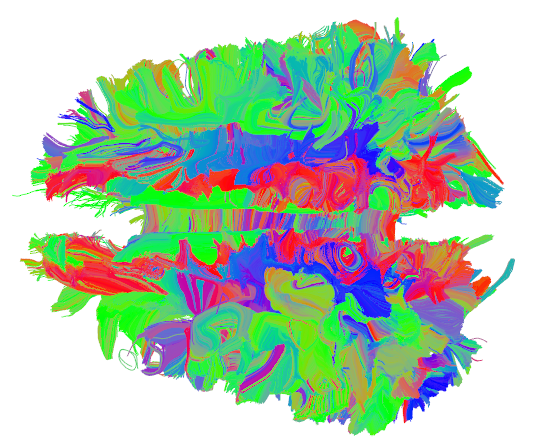}&
\includegraphics[width=0.24\textwidth,angle=270]{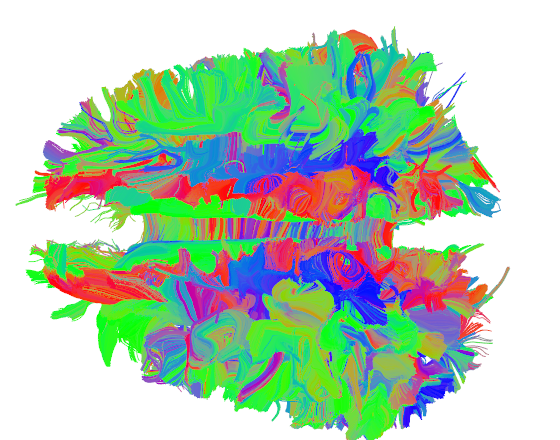}&
\includegraphics[width=0.24\textwidth,angle=270]{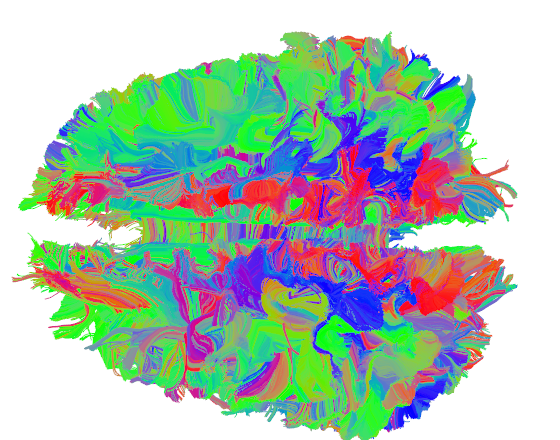}&
\includegraphics[width=0.24\textwidth,angle=270]{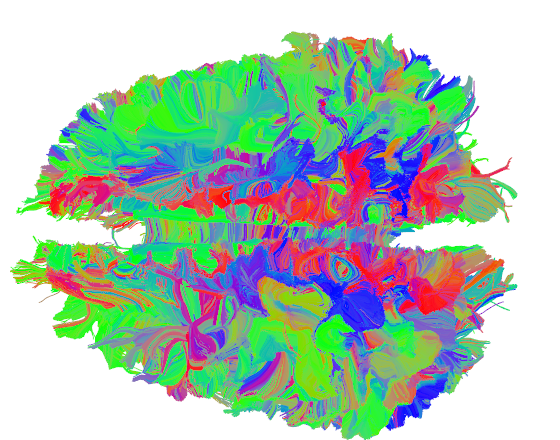}\\
\rotatebox{90}{\hspace{-2.0cm} 10 ($n=9$)}&
\includegraphics[width=0.24\textwidth,angle=270]{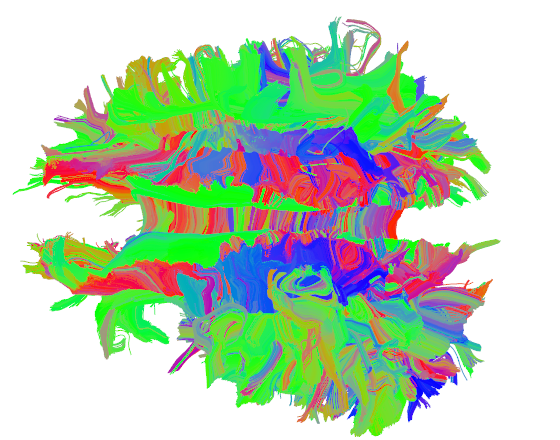}&
\includegraphics[width=0.24\textwidth,angle=270]{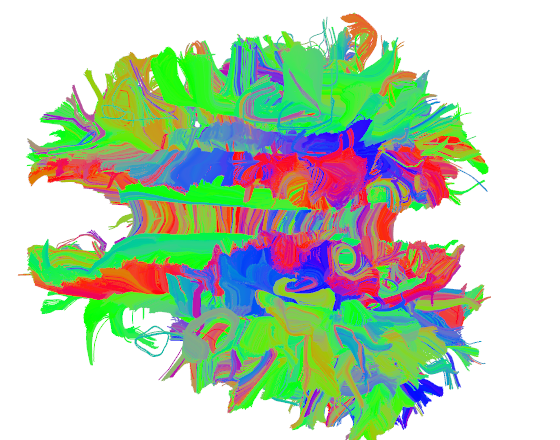}&
\includegraphics[width=0.24\textwidth,angle=270]{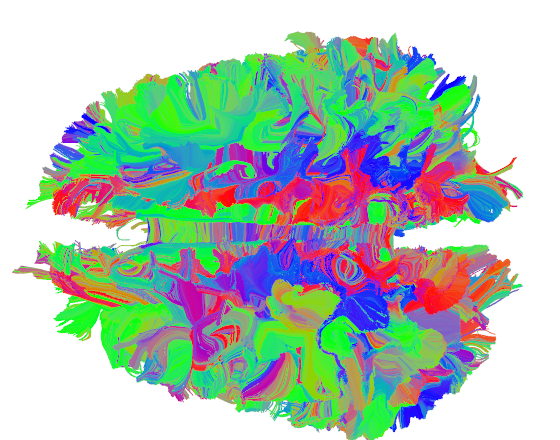}&
\includegraphics[width=0.24\textwidth,angle=270]{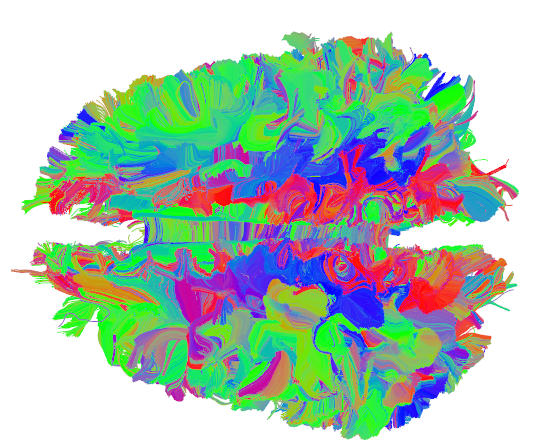}\\
\rotatebox{90}{\hspace{-2.0cm} 15 ($n=6$)}&
\includegraphics[width=0.24\textwidth,angle=270]{viz/15/fixed_sub_up.png}&
\includegraphics[width=0.24\textwidth,angle=270]{viz/15/learned_sub_up.png}&
\includegraphics[width=0.24\textwidth,angle=270]{viz/15/fixed_up.png}&
\includegraphics[width=0.24\textwidth,angle=270]{viz/15/learned_up.png}\\
\rotatebox{90}{\hspace{-2.0cm} 30 ($n=3$)}&
\includegraphics[width=0.24\textwidth,angle=270]{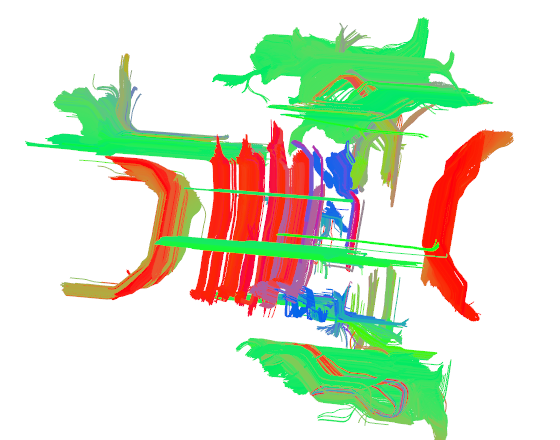}&
\includegraphics[width=0.24\textwidth,angle=270]{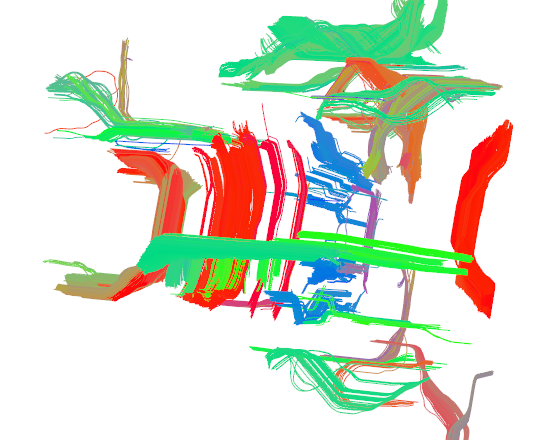}&
\includegraphics[width=0.24\textwidth,angle=270]{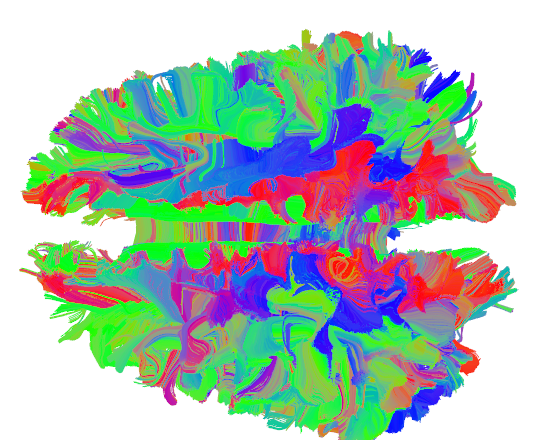}&
\includegraphics[width=0.24\textwidth,angle=270]{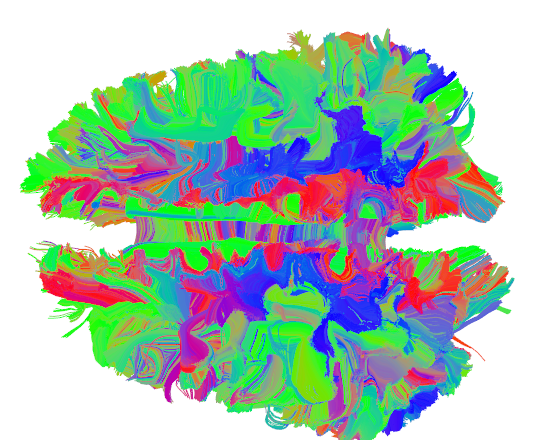}\\
\end{tabular}
    \caption{\textbf{Tractogram top-view.} Visual comparison of the full tractogram achieved from the DWI volume using fixed \& learned directions, with \& without reconstruction. Left-most column depicts the corresponding acceleration factor and the corresponding $n$ (number of diffusion directions).}
    \label{fig:supp-tract-up} 
\end{figure*}

\begin{figure*}
   \centering
   \addtolength{\tabcolsep}{-5.5pt}
\begin{tabular}{c c c c c}
AF (n)&Fixed w/ rec.&Learned-no rec.&Fixed-with rec&Learned-with rec.\\
\rotatebox{90}{\hspace{0.8cm} 3 ($n=30$)}&
\includegraphics[width=0.24\textwidth]{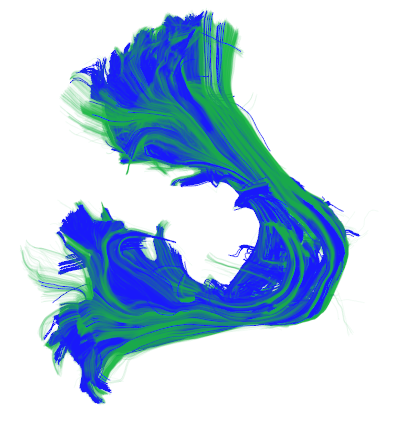}&
\includegraphics[width=0.24\textwidth]{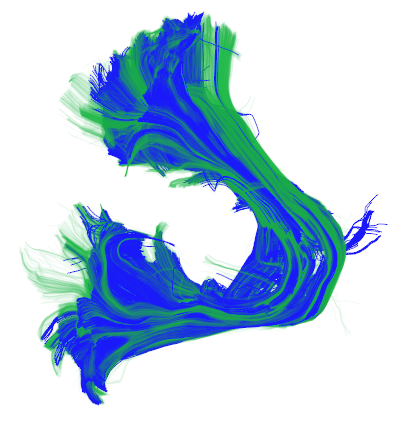}&
\includegraphics[width=0.24\textwidth]{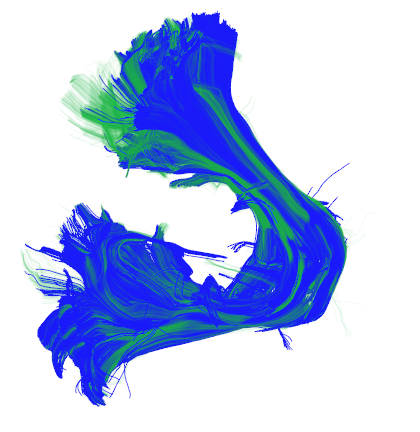}&
\includegraphics[width=0.24\textwidth]{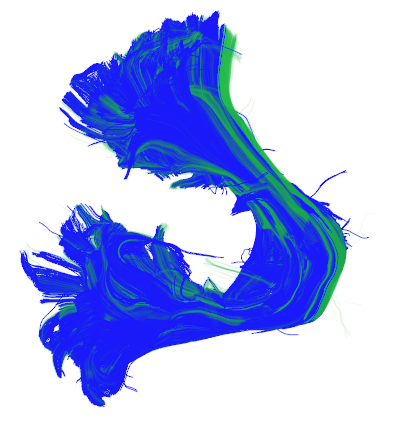}\\
\rotatebox{90}{\hspace{0.8cm} 5 ($n=18$)}&
\includegraphics[width=0.24\textwidth]{bundle/MCP/5fixed_sub.png}&
\includegraphics[width=0.24\textwidth]{bundle/MCP/5learned_sub.png}&
\includegraphics[width=0.24\textwidth]{bundle/MCP/5fixed.png}&
\includegraphics[width=0.24\textwidth]{bundle/MCP/5learned.png}\\
\rotatebox{90}{\hspace{0.8cm} 10 ($n=9$)}&
\includegraphics[width=0.24\textwidth]{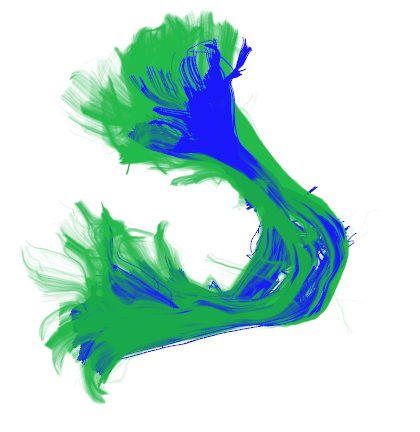}&
\includegraphics[width=0.24\textwidth]{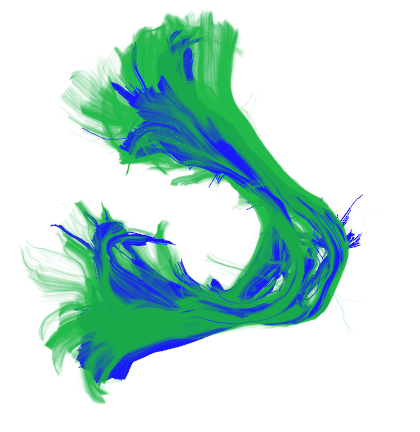}&
\includegraphics[width=0.24\textwidth]{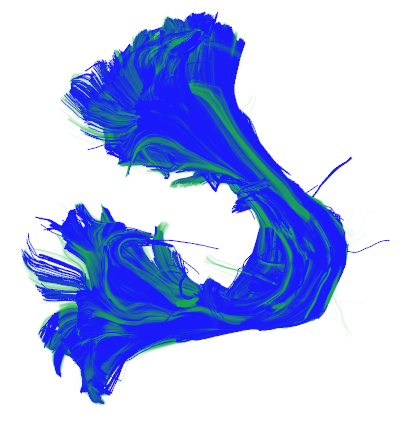}&
\includegraphics[width=0.24\textwidth]{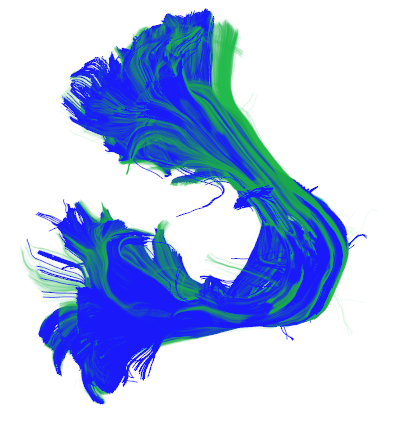}\\\\
\rotatebox{90}{\hspace{0.8cm} 15 ($n=6$)}&
\includegraphics[width=0.24\textwidth]{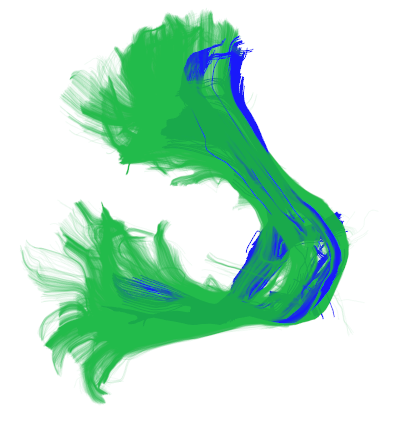}&
\includegraphics[width=0.24\textwidth]{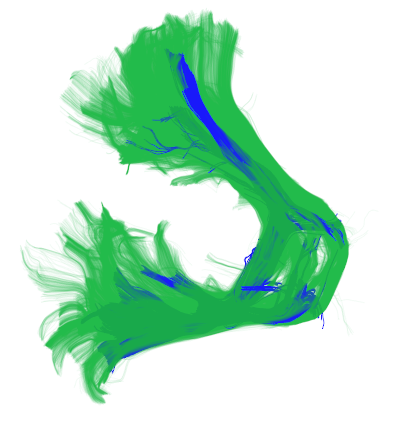}&
\includegraphics[width=0.24\textwidth]{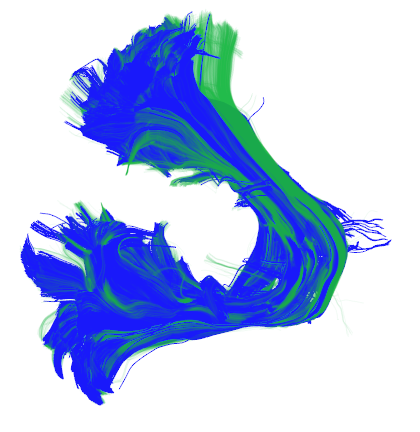}&
\includegraphics[width=0.24\textwidth]{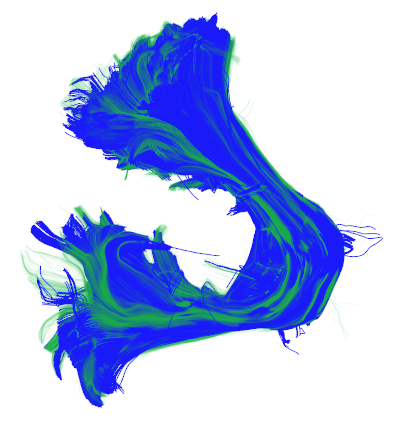}\\\\
\rotatebox{90}{\hspace{0.8cm} 30 ($n=3$)}&&&
\includegraphics[width=0.24\textwidth]{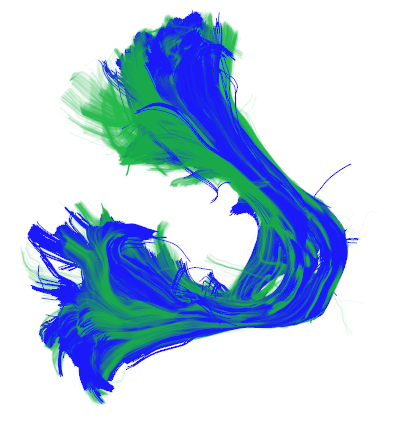}&
\includegraphics[width=0.24\textwidth]{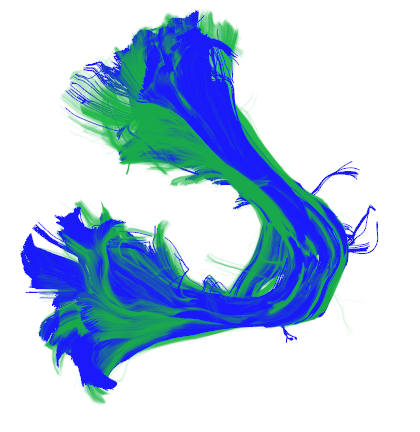}\\\\
\end{tabular}
    \caption{\textbf{Middle cerebellar peduncle bundle.} Visual comparison of the bundle achieved from applying tractography on the DWI volume using fixed \& learned directions with \& without reconstruction. Colored in green is the groundtruth bundle. Left-most column depicts the corresponding acceleration factor and the corresponding $n$ (number of diffusion directions).}
    \label{fig:supp-bundle-MCP} 
\end{figure*}

\begin{figure*}
   \centering
   \addtolength{\tabcolsep}{-5.5pt}
\begin{tabular}{c c c c c}
AF (n)&Fixed-no rec.&Learned-no rec.&Fixed-with rec&Learned-with rec.\\
\rotatebox{90}{\hspace{0.8cm} 3 ($n=30$)}&
\includegraphics[width=0.24\textwidth]{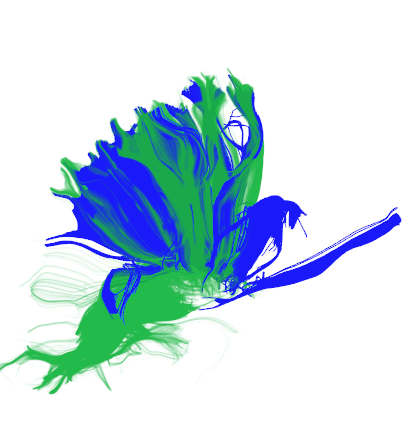}&
\includegraphics[width=0.24\textwidth]{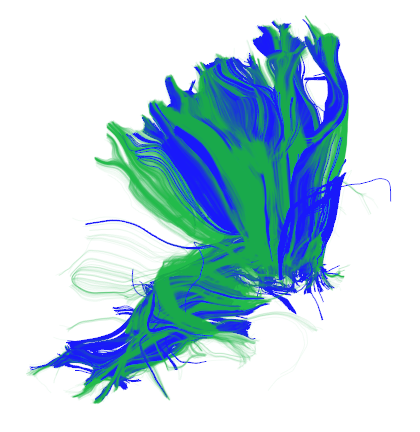}&
\includegraphics[width=0.24\textwidth]{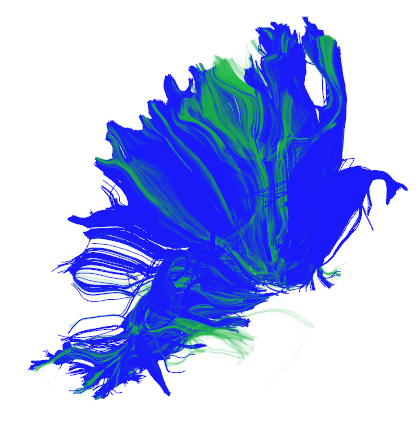}&
\includegraphics[width=0.24\textwidth]{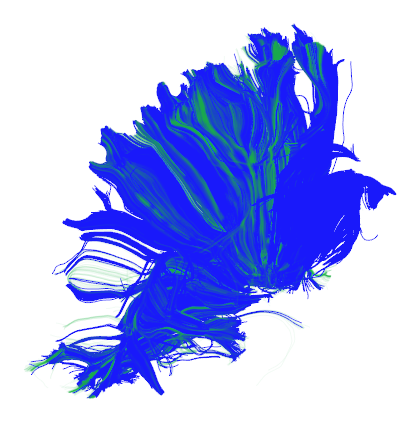}\\
\rotatebox{90}{\hspace{0.8cm} 5 ($n=18$)}&
\includegraphics[width=0.24\textwidth]{bundle/CS_L/5fixed_sub.png}&
\includegraphics[width=0.24\textwidth]{bundle/CS_L/5learned_sub.png}&
\includegraphics[width=0.24\textwidth]{bundle/CS_L/5fixed.png}&
\includegraphics[width=0.24\textwidth]{bundle/CS_L/5learned.png}\\
\rotatebox{90}{\hspace{0.8cm} 10 ($n=9$)}&
\includegraphics[width=0.24\textwidth]{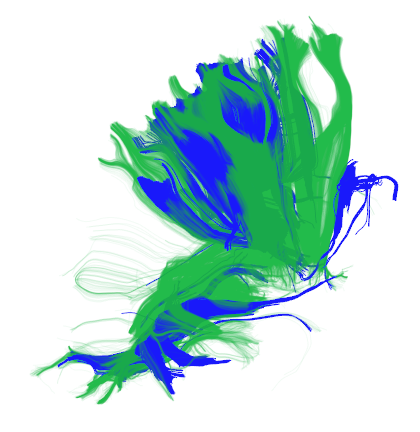}&
\includegraphics[width=0.24\textwidth]{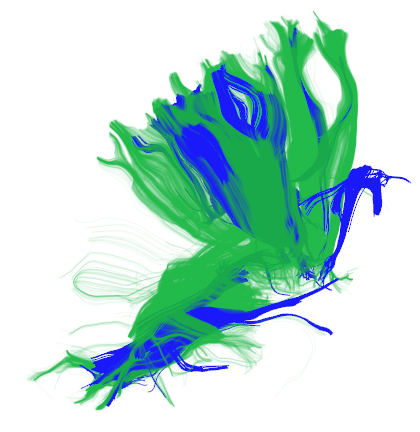}&
\includegraphics[width=0.24\textwidth]{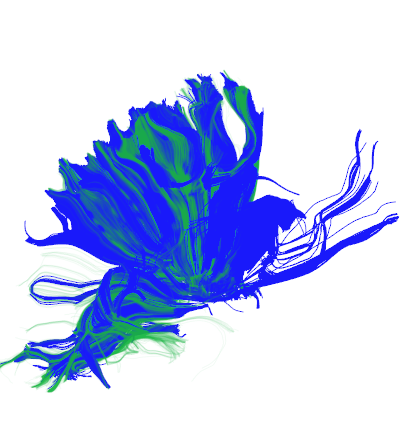}&
\includegraphics[width=0.24\textwidth]{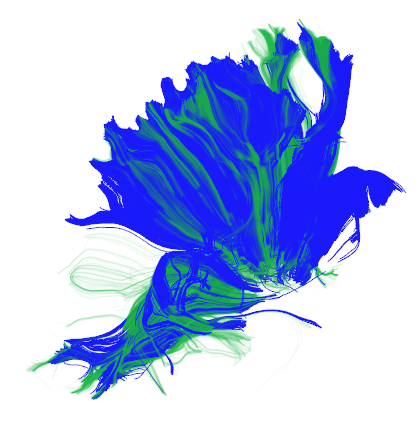}\\\\
\rotatebox{90}{\hspace{0.8cm} 15 ($n=6$)}&
\includegraphics[width=0.24\textwidth]{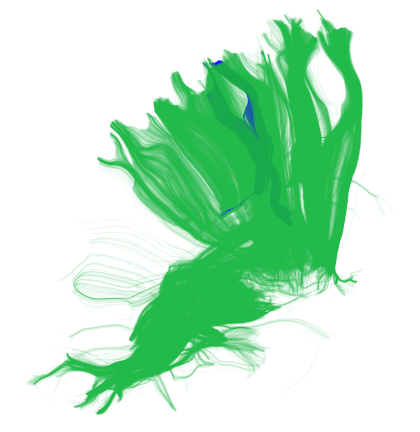}&
\includegraphics[width=0.24\textwidth]{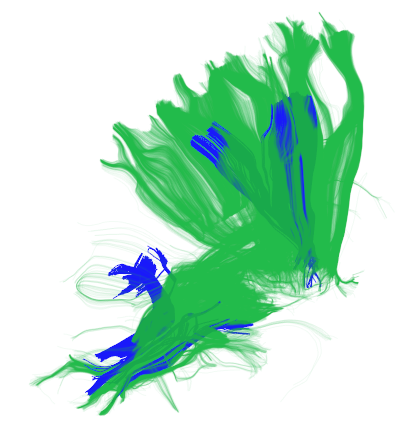}&
\includegraphics[width=0.24\textwidth]{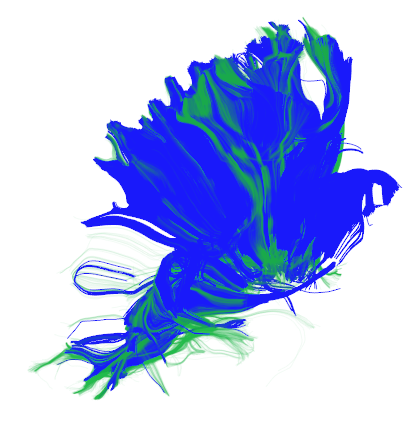}&
\includegraphics[width=0.24\textwidth]{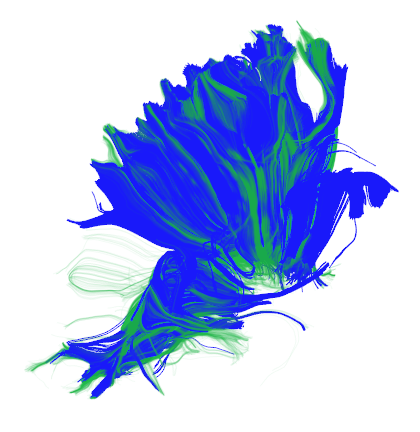}\\\\
\rotatebox{90}{\hspace{0.8cm} 30 ($n=3$)}&
\includegraphics[width=0.24\textwidth]{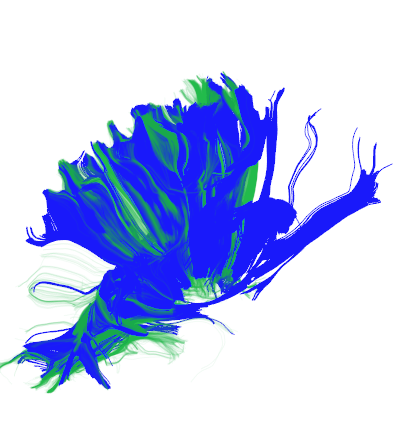}&
\includegraphics[width=0.24\textwidth]{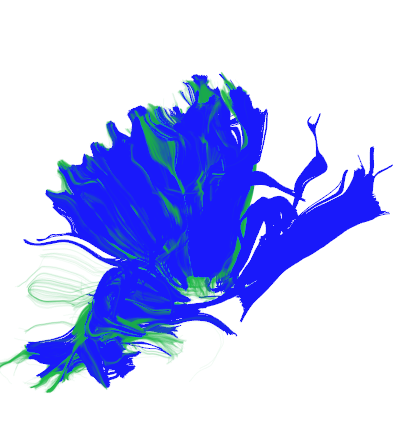}\\\\
\end{tabular}
    \caption{\textbf{CS Left bundle.}  Visual comparison of the bundle achieved from applying tractography on the DWI volume using fixed \& learned directions with \& without reconstruction. Colored in green is the groundtruth bundle. Left-most column depicts the corresponding acceleration factor and the corresponding $n$ (number of diffusion directions).}
    \label{fig:supp-bundle-CS_L} 
\end{figure*}

\end{document}